\documentclass[12pt,a4paper]{article}
\usepackage[T1]{fontenc}
\usepackage[dvipdfmx]{graphicx}
\usepackage{bm,caption,latexsym,amssymb,amsfonts,mathtools,here}
\usepackage{bm,latexsym,amsfonts,mathtools}
\usepackage{amssymb, amsmath, comment}
\usepackage{color}
\usepackage{hyperref}
\usepackage{booktabs}
\usepackage{mathrsfs}
\usepackage{wrapfig}
\usepackage{setspace}
\usepackage{braket}
\numberwithin{equation}{section}
\usepackage[sorting=none,doi=false]{biblatex}

\bibliography{reference}

\begin{document}
\begin{titlepage}
\unitlength = 1mm
\begin{flushright}
KUNS-3019,
KOBE-COSMO-24-04
\end{flushright}

\vskip 1cm
\begin{center}

{\large \textbf{
Imprints of Dark Photons on Gravitational Wave Polarizations
}}
\vspace{1.8cm}
\\
Kimihiro Nomura$^{\dagger}$, Jiro Soda$^{\flat}$, Kazushige Ueda$^{\%}$, and Ziwei Wang$^*$
\vspace{1cm}

\shortstack[l]
{\it $^\dagger$ Department of Physics, Kyoto University, Kyoto 606-8502, Japan}\\
{\it $^\flat$ Department of Physics, Kobe University, Kobe 657-8501, Japan\\
\it $^\%$
National Institute of Technology, Tokuyama College, Gakuendai, Shunan, Yamaguchi \\
\it 745-8585,Japan
\\
\it $^*$ Key Laboratory of Dark Matter and Space Astronomy,
Purple Mountain Observatory, \\
\it Chinese Academy of Sciences, Nanjing 210023, China  
}

\vskip 3.0cm

{\large Abstract}\\
\end{center}
We study conversion processes between gravitons and dark photons and reveal the effects of dark photons on the polarization of gravitational waves.
Considering cosmological dark magnetic fields, we investigate the evolution of the intensity and polarization of gravitational waves through the conversion. 
Specifically, we demonstrate that for minimal coupling between gravitons and dark photons, the intensity, circular polarization, and linear polarization evolve separately.
We derive explicit formulas for the statistical mean and variance of the intensity and polarization when the gravitational waves pass through magnetic fields with random orientation. The formulas capture how the initial polarization of dark photons will be imprinted on the observed gravitational wave background.

\vspace{1.0cm}
\end{titlepage}

\hrule height 0.075mm depth 0.075mm width 165mm
\tableofcontents
\vspace{1.0cm}
\hrule height 0.075mm depth 0.075mm width 165mm

\section{Introduction}
\label{sec:introduction}

Astronomy has historically developed by observations of electromagnetic waves.
We can gain various information about astrophysical objects from the intensity and the polarization of electromagnetic waves.
Remarkably, since the direct discovery of gravitational waves by LIGO in 2015 \cite{LIGOScientific:2016aoc}, gravitational wave astronomy has commenced \cite{Sesana:2016ljz,Sasaki:2018dmp}. Now and in the future, the intensity 
and the polarization of gravitational waves can be valuable tools to explore the universe.

It is known that the universe consists of visible matter, which is described by the Standard Model of particle physics, and invisible matter (dark matter), which can be controlled by physics beyond the Standard Model. Surprisingly, the majority of the matter of the universe belongs to the dark sector including the dark matter. 
Historically, the presence of dark matter has been suggested by observations of the visible matter affected by the dark matter through gravitational interactions. However, the dark sector can also directly affect the particles in the Standard Model such as photons.
For example, an axion field can interact with photons via the Chern-Simons coupling \cite{Kim:1986ax}.
We should note that magnetic fields exist in the universe on various scales, including stars \cite{Akiyama:2020dyx,Igoshev:2021ewx}, galaxies \cite{Widrow:2002ud,Jansson:2012pc}, intergalactic space \cite{HESS:2023zwb,Tjemsland:2023hmj,MAGIC:2022piy}, and so on \cite{Grasso:2000wj, Durrer:2013pga}. 
Remarkably, in the presence of magnetic fields, an axion particle can be converted into a photon with a polarization parallel to the magnetic fields~\cite{Raffelt:1987im, Maiani:1986md}. Since the axion can change one of the polarization states, this conversion process changes the polarization pattern of photons. In this way, the polarization pattern of photons carries information about the dark sector. 
Realistic magnetic fields will have some inhomogeneity.
Regarding this point, axion-photon conversion in random magnetic fields has been studied in Refs.~\cite{Grossman:2002by, Mirizzi:2009aj, Bassan:2010ya, Masaki:2017aea}.
Particularly, in Ref.~\cite{Masaki:2017aea}, the asymptotic behavior of Stokes parameters characterizing polarization of photons is investigated.
The results show that the equipartition of the degrees of freedom holds.

As mentioned above, on top of electromagnetic waves, gravitational waves can be useful to explore the universe.  
Furthermore, gravitational waves can convey information about the even earlier universe than electromagnetic waves.
Notice that it is natural to expect that the dark sector contains
dark gauge fields such as dark photons. Although dark photons with ultralight masses can be dark matter and generated during inflation~\cite{Dimopoulos:2006ms, Nakayama:2019rhg, Nakayama:2020rka, Nakai:2020cfw}, we should emphasize that dark photons are not necessarily the main component of dark matter. For instance, they may behave as dark radiation, mediating interactions in mirror dark matter model~\cite{Foot:1991kb, Foot:2004pa, Foot:2014mia}.
Now, we can raise the following question. What polarization pattern of primordial gravitational waves can be expected in the presence of dark photons? More specifically, since dark photons are analogous to the Standard Model photons, it is legitimate to assume cosmological dark magnetic fields~\cite{Masaki:2018eut}. In that case, dark photons can be converted into gravitons and vice versa due to the universal (minimal) coupling~\cite{Gertsenshtein:1962, Raffelt:1987im}. 
(Here, ``gravitons'' usually stand for quanta corresponding to gravitational waves, but the conversion itself can also be derived from the classical description.)
Hence, the polarization pattern can be affected by dark photons.
In this paper, we consider graviton - dark photon conversion in cosmological dark magnetic fields and reveal the evolution of the Stokes parameters of gravitational waves.

This paper is organized as follows.
In Section \ref{sec:conversion}, we study the conversion between gravitational waves and dark photons in a domain of a cosmological dark magnetic field where the magnetic field is spatially uniform.
In Section \ref{sec:network}, after constructing a sequence of domains of magnetic fields with random directions, we derive equations describing the evolution of the Stokes parameters of gravitational waves affected by conversion through the domains.
In Section \ref{sec:imprints}, we investigate the asymptotic behavior of the Stokes parameters when gravitational waves pass through many random magnetic field domains. In particular, we examine the statistical mean and variance of the intensity, circular polarization, and linear polarization of gravitational waves.
The final section is devoted to the conclusion and discussion.
We adopt the natural units, $c = \hbar = \epsilon_0 = 1$ where $c$ is the speed of light, $\hbar$ is the reduced Planck constant, and $\epsilon_0$ is the permittivity in a vacuum. The metric signature is $(-,+,+,+)$.

\section{Cosmological dark magnetic fields and graviton - dark photon conversion}
\label{sec:conversion}

In this paper, we consider gravitational waves and waves of a dark $U(1)$ gauge field propagating in the expanding universe accompanied by cosmological dark magnetic fields. 

We start with the Einstein--Hilbert action and the action of the $U(1)$ gauge field,
\begin{align}
S=\int d^4x \, \sqrt{-g}
\left[
\frac{1}{2\kappa^2} R
-\frac{1}{4}g^{\mu\rho} g^{\nu\sigma}
F_{\mu\nu} F_{\rho\sigma}
\right]. 
\label{eqaction1}
\end{align}
Here, $R$ represents the Ricci scalar associated with the metric $g_{\mu\nu}$, $g$ is the determinant of $g_{\mu\nu}$,  
$F_{\mu\nu} = \partial_\mu A_\nu - \partial_\nu A_\mu$ is the field strength tensor of the $U(1)$ gauge field $A_\mu$, and
$\kappa^2 \equiv 8\pi G \equiv M_{\text{Pl}}^{-2}$, where $G$ is the Newton constant and $M_{\text{Pl}}$ is the reduced Planck mass.
The variation of the action with respect to the metric gives the Einstein equation as
\begin{align}
{G^{\mu}}_{\nu}=\kappa^2 \left(F^{\mu\alpha} F_{\nu\alpha}-\frac{1}{4}\delta^{\mu}_{\nu}F_{\alpha\beta}F^{\alpha\beta}\right)\,,
\label{EinsteinEQ}
\end{align}
where $G_{\mu\nu}$ is the Einstein tensor.
The variation with respect to the gauge field $A_\mu$ gives the Maxwell equation as
\begin{align}
\partial_\mu \left( \sqrt{-g}\, F^{\mu\nu}\right) =0\,.
\label{MaxwellEQ}
\end{align}

\subsection{Dark magnetic fields}

The background spacetime is described by the spatially flat Friedmann-Lema\^itre-Robertson-Walker metric, 
\begin{align}
    \bar{g}_{\mu\nu} dx^\mu dx^\nu 
    = a^2 ( -d\eta^2 + \delta_{ij} \, dx^i dx^j )\,, 
    \label{eqmet1-1}
\end{align}
where $\eta (=x^0)$ is the conformal time, $x^i$ with $i = 1,2,3$ are the comoving coordinates, and $a=a(\eta)$ is the scale factor.

To consider background magnetic fields and propagating waves of the $U(1)$ gauge field, we write the total gauge field strength as
\begin{align}
    F_{\mu\nu} = \bar{F}_{\mu\nu} + \partial_\mu A_\nu - \partial_\nu A_\mu\,. 
    \label{eqgau1}
\end{align}
Here, $\bar{F}_{\mu\nu}$ denotes the background field strength, and $A_\mu$ denotes the propagating field on that background, which is treated as a perturbation.
(Note the change in notation: At the beginning of this section, $A_{\mu}$ denoted the total gauge field, but hereafter $A_{\mu}$ stands for the perturbation.)
For the gauge field perturbation $A_\mu$, we take the radiation gauge 
\begin{align}
    \partial_i A_i = 0 \,,
    \qquad A_0 = 0\,. 
    \label{eqgau2}
\end{align}

Since we consider the magnetic configuration as the background, we assume $\bar{F}_{0i} = 0$ in the coordinate system $\{\eta, \bm{x}(=x^i) \}$.
In this case, the $\mu = i$ component of the Bianchi identity $\epsilon^{\mu\nu\rho\sigma} \partial_{\nu} \bar{F}_{\rho\sigma} = 0$ tells us that $\bar{F}_{ij}$ is constant in time. 
Then, we see that the energy density of the background dark magnetic field behaves as 
\begin{align}
    \frac{1}{4} \bar{F}^{ij} \bar{F}_{ij} \propto a^{-4}\,, 
    \label{eqB1-1}
\end{align}
by noting that the spatial indices of $\bar{F}^{ij}$ are raised by the background inverse metric $\bar{g}^{ij} = a^{-2} \delta_{ij}$.
Let us introduce the physical dark magnetic field vector $\bar{\bm{B}}(a) = (\bar{B}_i(a))$ as 
\begin{align}
    \bar{B}_i(a) \equiv \frac{1}{2\, a^2} \epsilon_{ijk} \bar{F}_{jk} \,, 
    \label{eqB1}
\end{align}
where $\epsilon_{ijk}$ is the anti-symmetric symbol normalized as $\epsilon_{123} = 1$.
The vector $\bar{\bm{B}}(a)$ is defined so that the magnetic field energy density is represented as 
\begin{align}
    \frac{1}{4} \bar{F}^{ij} \bar{F}_{ij} = \frac{1}{2} \bar{B}^2(a) \,,
    \label{eqB2}
\end{align}
where $\bar{B}^2(a) \equiv \delta_{ij} \bar{B}_i(a) \bar{B}_j(a)$.
Denoting the physical magnetic field magnitude at present by $\bar{B}_0$, the magnitude at the scale factor $a$ is given by
\begin{align}
    \bar{B}(a) = \frac{\bar{B}_0}{a^2}\,, 
    \label{eqB3}
\end{align}
where we normalized the present scale factor to be unity.
Throughout this paper, we assume that the magnetic field would be rooted in a primordial origin, and the physical magnitude is only diluted through the cosmic expansion as Eq.~\eqref{eqB3}.

If we identify $\bar{B}_0$ with the present magnetic field made of the Standard Model (SM) photons, various observations tell us a constraint 
\begin{align}
    10^{-16}~\text{Gauss} \lesssim \bar{B}_0 \lesssim 10^{-9}~\text{Gauss}\,.
    \qquad \text{(SM photon)} 
    \label{eqB4}
\end{align}
The lower limit is suggested by gamma-ray observations \cite{Neronov:2010gir, Taylor:2011bn, Xia:2022uua}, and the upper limit comes from Cosmic Microwave Background (CMB) and matter power spectrum observations \cite{Yamazaki:2012pg}.
On the other hand, when we consider the magnetic field as made of a $U(1)$ gauge field in a dark sector that does not interact with the SM, the upper bound will be relaxed.
In this case, the constraint would be inferred from Big Bang Nucleosynthesis (BBN).
We employ the upper limit of the dark magnetic field magnitude obtained in the literature \cite{Kawasaki:2012va},
\begin{align}
    \bar{B}_0 \lesssim 10^{-6} ~ \text{Gauss}\,. 
    \qquad \text{(dark photon)}
    \label{constraint2}
\end{align}

Below, we assume that the background magnetic field is spatially uniform, $\partial_i \bar{B}_j = 0$, in a certain domain of interest. 
We expect that the model with a uniform background magnetic field works well if the coherence length of the magnetic field is sufficiently longer than the wavelength of the gravitational and gauge field waves.

\subsection{Graviton - dark photon conversion}

To incorporate gravitational waves, we write the total metric as
\begin{align}
    g_{\mu\nu} dx^\mu dx^\nu 
    = a^2 \big[ -d\eta^2 +  (\delta_{ij} + h_{ij}) \, dx^i dx^j  \big]\,, 
    \label{eqmet1}
\end{align}
where $h_{ij} = h_{ij}(\eta, \bm{x})$ represents the metric perturbation propagating as gravitational waves.
For $h_{ij}$, we take the transverse-traceless gauge, 
\begin{align} 
    h_{ii} = 0 \,, 
    \qquad \partial_{j} h_{ij} = 0 \,. 
    \label{eqmet2}
\end{align}
 
Substituting Eqs.~\eqref{eqgau1} and \eqref{eqmet1} into the Einstein equation \eqref{EinsteinEQ}, and extracting the linear terms with respect to $h_{ij}$ and $A_i$, we obtain the equation of motion for gravitational waves as 
\begin{align}
    (\partial_\eta^2 + 2\mathcal{H} \partial_\eta - \nabla^2) h_{ij} 
    = 2 \kappa^2 \, [ \epsilon_{ikl} \bar{B}_l(a) (\partial_j A_k - \partial_k A_j) + \epsilon_{jkl} \bar{B}_l(a) (\partial_i A_k - \partial_k A_i) ]\,,
    \label{eqh1}
\end{align}
where $\mathcal{H} \equiv (\partial_{\eta} a) / a$, $\nabla^2 \equiv \delta_{ij} \partial_i \partial_j$, and the vector $\bar{B}_i(a)$ is defined in Eq.~\eqref{eqB1}. 
Equation \eqref{eqh1} indicates that, in the presence of the background magnetic field $\bar{B}_i$, the gauge fields $A_i$ on the right-hand side serve as the source of gravitational waves $h_{ij}$.
Note that we have neglected the terms of the form of $a^2 \kappa^2 \bar{B}^2 h$ on the right-hand side. 
The contribution of these terms is suppressed by the factor $a \, \bar{B} / (k\, M_{\text{Pl}})$ compared with that of $\kappa^2\bar{B} (\partial A)$ terms, where $k$ abstractly represents the comoving wavenumber of the perturbation.
Throughout this paper, we consider the perturbations with large wavenumber $k$ so that we can neglect the $a^2 \kappa^2\bar{B}^2 h$ terms.\footnote{
Using the scaling $\bar{B} = \bar{B}_0 / a^2$ from Eq.~\eqref{eqB3}, we have the expression $a \bar{B} / (k M_{\text{Pl}}) \sim 10^{-22} a^{-1} ( \bar{B}_0 / 10^{-6} \, \text{Gauss} ) ( 100\,\text{Hz} / k )$.
This means that for a magnetic field amplitude near the upper bound of $B_0 \sim 10^{-6}\, \text{Gauss}$ (see Eq.~\eqref{constraint2}) and a wavenumber within the LIGO band, such as $k \sim 100 \, \text{Hz}$, the $a^2\kappa^2 \bar{B}^2 h$ terms can be disregarded. This is a valid approximation when $a \gg 10^{-22}$. Consequently, Eq.~\eqref{eqh1} is safely applicable even during the radiation-dominated epoch.
}

Similarly, from the $\nu = i$ component of the Maxwell equation \eqref{MaxwellEQ}, we obtain the linear order equation of motion for the gauge field as
\begin{align}
(\partial_\eta^2 - \nabla^2) A_i = a^2 \epsilon_{jkl} \bar{B}_{l}(a)\, \partial_k h_{ij}\,.
\label{eqA1}
\end{align}
This equation shows that gravitational waves $h_{ij}$ can be the source of gauge fields $A_i$ in the presence of the magnetic field $\bar{B}_i$. Combining Eqs.~\eqref{eqh1} and \eqref{eqA1}, we see that gravitational waves and gauge fields are converted into each other in the magnetic field background. In the rest of this section, we solve the above equations and explicitly show that the conversion occurs. The conversion was originally discussed in \cite{Gertsenshtein:1962, Raffelt:1987im}. The present treatment is an extension to the cosmological context \cite{Masaki:2018eut}.

The gauge field $A_i$ is treated as a massless field so far. 
However, for the Standard Model (SM) photons, an effective mass is induced by the effects of non-vacuum environments, such as plasma oscillations and quantum corrections in large magnetic fields. 
For dark photons that are not part of the SM, possibly they have an intrinsic mass.
Such (effective) masses for gauge fields generally suppress conversion with gravitational waves. 
Since our main concern is the cases such that conversion is efficient, hereafter we neglect the (effective) masses for simplicity. 
For dark photons, this treatment is valid at any frequency if they are decoupled with other particles and are massless.

As an alternative to $h_{ij}$, it is convenient to introduce a new variable $y_{ij}$ as  
\begin{align}
    y_{ij} \equiv (2\kappa)^{-1} \,a \, h_{ij}\,. 
    \label{eqh1-2}
\end{align}
Then, the equations of motion \eqref{eqh1} and \eqref{eqA1} are recast to 
\begin{align}
    \bigg( \partial_\eta^2 - \nabla^2 - \frac{\partial_\eta^2 a}{a} \bigg) y_{ij} 
    &= \kappa \,a \, [ \epsilon_{ikl} \bar{B}_l(a)\, (\partial_j A_k - \partial_k A_j) + \epsilon_{jkl} \bar{B}_l(a)\, (\partial_i A_k - \partial_k A_i) ] \,, 
    \label{eqh2-2}
    \\
    (\partial_\eta^2 - \nabla^2) A_i &= 2 \,\kappa \,a \, \epsilon_{jkl} \bar{B}_{l}(a)\, \partial_k y_{ij} \,.
    \label{eqA2-2}
\end{align}

We define the Fourier decomposition for the metric and gauge field perturbations as 
\begin{align}
&{y}_{ij}(\bm{x},\eta)
=
\sum_{P=+,\times}
\frac{1}{(2\pi)^{3/2}}
\int d^3{\bm k}\,
{y}_P({\bm k},\eta) \,
e^P_{ij}(\hat{\bm{k}}) \,
e^{i{\bm k}\cdot {\bm x}},
\label{polRELATION2}\\
&{A}_i(\bm{x}, \eta) =
\sum_{P=+,\times}
\frac{i}{(2\pi)^{3/2}}
\int d^3{\bm k}\,
{A}_P({\bm k},\eta) \,
e^P_i(\hat{\bm{k}}) \,
e^{i{\bm k}\cdot{\bm x}},
\label{polRELATION}
\end{align}
where we defined the unit vector $\hat{\bm{k}}\equiv {\bm k}/k$ with $k \equiv |\bm{k}| \equiv \sqrt{\delta_{ij} k^i k^j}$.
For each direction $\hat{\bm{k}}$, we introduced the polarization vectors $e^{P}_{i} (\hat{\bm{k}})$ and the polarization tensors $e^{P}_{ij} (\hat{\bm{k}})$ with $P = +, \times$ for the transverse components.
The polarization vectors are introduced to be real and satisfy 
\begin{align}
    &k^i \, e^P_{i}(\hat{\bm{k}}) = 0 \,, 
    \label{eqe1} \\
    &e_i^P(\hat{\bm{k}}) \, e_i^Q(\hat{\bm{k}}) = \delta^{PQ} \,,
    \label{eqe2} \\ 
    &e^+_i(\hat{\bm{k}}) = - e^+_i(-\hat{\bm{k}}) \,, \qquad
    e^\times_i (\hat{\bm{k}}) = e^\times_i (-\hat{\bm{k}})\,. 
    \label{eqe3}
\end{align}
Given $e^P_{i}(\hat{\bm{k}})$, we construct the polarization tensors $e^{P}_{ij} (\hat{\bm{k}})$ as 
\begin{align}
    e^+_{ij}(\hat{\bm{k}})
    &= \frac{1}{\sqrt{2}}
    \left[e^+_i (\hat{\bm{k}}) \, e^+_j (\hat{\bm{k}}) - e^\times_i (\hat{\bm{k}}) \, e^\times_j (\hat{\bm{k}})\right] \,, 
    \label{eqe4} \\
    e^\times_{ij}(\hat{\bm{k}})
    &= \frac{1}{\sqrt{2}}
    \left[e^+_i (\hat{\bm{k}}) \, e^\times_j (\hat{\bm{k}}) + e^\times_i (\hat{\bm{k}}) \, e^+_j (\hat{\bm{k}})
    \right]\,. 
    \label{eqe5}
\end{align}
These tensors satisfy 
\begin{align}
    &k^i \, e^P_{ij}(\hat{\bm{k}}) = 0 \, , \qquad 
    e^P_{ii}(\hat{\bm{k}}) = 0 \, , 
    \label{eqe6} \\
    &e_{ij}^P (\hat{\bm{k}}) \, e_{ij}^Q(\hat{\bm{k}})=\delta^{PQ} \, , 
    \label{eqe7} \\
    &e^+_{ij}(\hat{\bm{k}}) = e^+_{ij}(-\hat{\bm{k}}) \,, \qquad 
    e^\times_{ij}(\hat{\bm{k}}) = -e^\times_{ij}(-\hat{\bm{k}}) \,. 
    \label{eqe8}
\end{align}
Since the fields $h_{ij}(\bm{x}, \eta)$ and $A_i(\bm{x}, \eta)$ are real (Hermite), the Fourier components $y_{P} (\bm{k}, \eta)$ and $A_{P} (\bm{k}, \eta)$ must satisfy 
\begin{align}
    &y_{+} (\bm{k}, \eta) = y_{+}^* (-\bm{k}, \eta) \, , \qquad 
    y_{\times} (\bm{k}, \eta) = - y_{\times}^* (-\bm{k}, \eta) \,,
    \label{herm1}
    \\
    &A_{+} (\bm{k}, \eta) = A_{+}^* (-\bm{k}, \eta) \, , \qquad 
    A_{\times} (\bm{k}, \eta) = -A_{\times}^* (-\bm{k}, \eta) \,,
    \label{herm2}
\end{align}
where the asterisk (*) denotes the complex conjugate.

We define the polarization vectors so that the magnetic field vector $\bar{\bm{B}}(a)$ lies in the plane spanned by $\hat{\bm{k}}$ and $\bm{e}^\times(\hat{\bm{k}})$ without loss of generality. 
Using $\{ \bm{e}^+(\hat{\bm{k}}), \bm{e}^\times(\hat{\bm{k}}), \hat{\bm{k}} \}$ as the coordinate axes for a given $\bm{k}$ and introducing the polar angle $\theta$, we represent the background magnetic field vector $\bar{\bm{B}}(a)$ as
\begin{align}
    \bar{\bm{B}}(a) = ( 0 , \bar{B}(a) \sin \theta , \bar{B}(a) \cos \theta )\,, 
    \label{eqB5}
\end{align}
in that coordinate system.
Inserting the Fourier decomposition \eqref{polRELATION2} and \eqref{polRELATION} into Eqs.~\eqref{eqh2-2} and \eqref{eqA2-2}, and multiplying polarization vectors and tensors, we obtain the equation of motion for each polarization $P = +, \times$ as
\begin{align}
    \left[ \partial_\eta^2 + k^2 +
    \begin{pmatrix}
    -(\partial_\eta^2 a)/a &0 \\
    0    &0
    \end{pmatrix}
    \right]
    \begin{pmatrix}
        y_P (\bm{k},\eta) \\
        A_P (\bm{k},\eta)
    \end{pmatrix}
    = - k\,\lambda(a)
    \begin{pmatrix}
        0   &1  \\
        1   &0  
    \end{pmatrix}
    \begin{pmatrix}
        y_P (\bm{k},\eta)\\
        A_P (\bm{k},\eta)
    \end{pmatrix} ,
    \label{eqhA1}
\end{align}
where we defined 
\begin{align}
    \lambda(a) \equiv \sqrt{2} \, \kappa \, a \, \bar{B}(a) \sin \theta = \sqrt{2} \, \kappa \, \frac{\bar{B}_0}{a} \sin \theta \,. 
    \label{eqhA2}
\end{align}
Equation \eqref{eqhA1} has been derived in Ref.~\cite{Kanno:2023fml}, where it was applied to the inflationary universe and it was shown that tachyonic growth of primordial gravitational waves may appear.

From Eq.~\eqref{eqhA1}, we see that gravitons of the $+/\times$ mode have mixing with gauge fields of the $+/\times$ mode. 
Let us now investigate conversion between gravitons and gauge fields.
To see conversion, it is useful to write the components $y_P(\bm{k}, \eta)$ and $A_P(\bm{k}, \eta)$ as 
\begin{align}
    y_+(\bm{k}, \eta) &= \tilde{y}_+ (\bm{k}, \eta) \, e^{-ik\eta} + \tilde{y}_+^* (-\bm{k}, \eta) \, e^{+ik\eta} \,,
    \label{eqh3}
    \\
    y_\times(\bm{k}, \eta) &= \tilde{y}_\times (\bm{k}, \eta) \, e^{-ik\eta} - \tilde{y}_\times^* (-\bm{k}, \eta) \, e^{+ik\eta} \,,
    \label{eqh4}
    \\
    A_+(\bm{k}, \eta) &= \tilde{A}_+ (\bm{k}, \eta) \, e^{-ik\eta} + \tilde{A}_+^* (-\bm{k}, \eta) \, e^{+ik\eta} \,,
    \label{eqA3}
    \\
    A_\times(\bm{k}, \eta) &= \tilde{A}_\times (\bm{k}, \eta) \, e^{-ik\eta} - \tilde{A}_\times^* (-\bm{k}, \eta) \, e^{+ik\eta} \,.
    \label{eqA4}
\end{align}
(In each equation, the sign of the second term on the right-hand side inherits the properties  \eqref{herm1} and \eqref{herm2}.)
Inserting Eqs.~\eqref{eqh3}--\eqref{eqA4} into Eq.~\eqref{eqhA1} and assuming $|\partial_\eta^2 \tilde{y}_P| \ll k |\partial_\eta \tilde{y}_P|$ and $|\partial_\eta^2 \tilde{A}_P| \ll k |\partial_\eta \tilde{A}_P|$, we obtain a Schr\"odinger-like equation  
\begin{align}
    i \, \partial_\eta \vec{\Psi}_P
    = \bm{H}(a) \,\vec{\Psi}_P 
    \label{eqsch1}
\end{align}
with 
\begin{align}
    \vec{\Psi}_P \equiv
    \begin{pmatrix}
        \tilde{y}_P(\bm{k}, \eta) \\
        \tilde{A}_P(\bm{k}, \eta)
    \end{pmatrix}\,,
    \qquad 
    \bm{H}(a) \equiv 
    \begin{pmatrix}
        \Delta_h(a)    &\Delta_{M}(a)  \\
        \Delta_{M}(a)  &0
    \end{pmatrix}\,, 
    \label{eqsch2}
\end{align}
where we defined 
\begin{align}
    \Delta_{M}(a) &\equiv \frac{\lambda(a)}{2} = \frac{\kappa}{\sqrt{2}} \frac{\bar{B}_0}{a} \sin \theta \,,
    \label{deltaM}
    \\
    \Delta_h(a) &\equiv - \frac{1}{2k} \frac{\partial_\eta^2 a}{a} \,.
    \label{deltah}
\end{align}

Since we are interested in cases of efficient conversion, hereafter we assume 
\begin{align}
    |\Delta_{M}(a)| \gg |\Delta_h(a)|\, , 
    \label{eqsch3}
\end{align}
so that we neglect the diagonal component of $\bm{H}(a)$.
Particularly, in the radiation-dominated epoch in which $a \propto \eta$, the component $\Delta_h(a)$ vanishes. 
More generally, we can estimate as $|\Delta_h(a)| \sim \mathcal{H}^2 / k$ omitting some numerical factor, where $\mathcal{H} = (\partial_\eta a) / a$ is the comoving Hubble scale, and we are interested only in large $k$ so that $|\Delta_h(a)|$ is negligible.

Once the diagonal component is neglected, the matrix $\bm{H}(a)$ can be diagonalized as 
\begin{align}
    \bm{R}^\top \bm{H}(a) \bm{R} = 
    \begin{pmatrix}
        + \Delta_{M}(a)  &0 \\
        0    &- \Delta_{M}(a)
    \end{pmatrix}\,,
    \qquad 
    \bm{R} = \frac{1}{\sqrt{2}}
    \begin{pmatrix}
        1 &-1 \\
        1 &1
    \end{pmatrix}\,. 
    \label{eqsch4}
\end{align}
The important notice is that $\bm{R}$ is a constant orthogonal matrix. 
Hence, we can obtain the analytic solution as 
\begin{align}
    \vec{\Psi}_P(\eta) 
    = \bm{R} 
    \begin{pmatrix}
        e^{-i\int_{\eta_\text{i}}^\eta d\eta \, \Delta_M (a) }   &0 \\
        0   &e^{+i\int_{\eta_\text{i}}^\eta d\eta \, \Delta_M (a) }
    \end{pmatrix} 
    \bm{R}^\top \vec{\Psi}_P (\eta_{\text{i}}) \,, 
    \label{eqsch5}
\end{align}
where $\vec{\Psi}_P (\eta_{\text{i}})$ is specified by the initial condition at $\eta = \eta_{\text{i}}$.
Finally, the time evolution of the variables $\tilde{y}_P(\bm{k},\eta)$ and $\tilde{A}_P(\bm{k},\eta)$ reads
\begin{align}
    \tilde{y}_P (\bm{k},\eta) 
    &= \tilde{y}_P (\bm{k},\eta_{\text{i}})  \cos \bigg( \int_{\eta_{\text{i}}}^\eta d\eta \, \Delta_M (a) \bigg) \, 
    - i \,\tilde{A}_P (\bm{k},\eta_{\text{i}}) \sin \bigg( \int_{\eta_{\text{i}}}^\eta d\eta \, \Delta_M (a) \bigg) \,,
    \label{ysol1}
    \\
    \tilde{A}_P (\bm{k},\eta) 
    &=\tilde{A}_P (\bm{k},\eta_{\text{i}}) \cos \bigg( \int_{\eta_{\text{i}}}^\eta d\eta \, \Delta_M (a) \bigg) \, 
    - i \,\tilde{y}_P(\bm{k},\eta_{\text{i}})  \sin \bigg( \int_{\eta_{\text{i}}}^\eta d\eta \, \Delta_M (a) \bigg)  \,.
    \label{Asol1}
\end{align}

The $\tilde{y}_P(\bm{k}, \eta)$ and $\tilde{A}_P(\bm{k}, \eta)$ correspond to the (complex) amplitudes of the gravitational wave and gauge field with wavenumber vector $\bm{k}$ at time  $\eta$, respectively. 
For example, given the initial condition $\tilde{y}_P(\bm{k}, \eta_{\text{i}}) = 0$, the conversion probability from the gauge fields to the gravitons $P_{\gamma \to h}$ is given by
    \begin{align}
    P_{\gamma \to h}(\eta) 
    \equiv \frac{|\tilde{y}_P(\bm{k}, \eta)|^2}{|\tilde{A}_P(\bm{k}, \eta_{\text{i}})|^2}
    = \sin^2 \bigg( \int_{\eta_{\text{i}}}^\eta d\eta \, \Delta_M(a) \bigg), 
    \end{align} 
which depends on time $\eta$, or equivalently, the propagation distance.

We see that conversion between gravitons and gauge fields occurs on the length scale $\Delta^{-1}_{M}(a)$, which is numerically given by
\begin{align}
    \Delta^{-1}_{M}(a) = 1.1 \times 10^{6} \, a~\text{Mpc} 
    \left( \frac{10^{-6}~\text{Gauss}}{\bar{B}_0 \sin \theta} \right) \,.
    \label{length}
\end{align}

In the above derivation, we have assumed $|\partial_\eta^2 \tilde{y}_P| \ll k |\partial_\eta \tilde{y}_P|$ and $|\partial_\eta^2 \tilde{A}_P| \ll k |\partial_\eta \tilde{A}_P|$.
For consistency of the solution, the conditions $k \gg |\Delta_M|, |\partial_\eta \Delta_M|/|\Delta_M|$ should be satisfied, which are translated into\footnote{
Using the estimate of $\Delta_M$ in Eq.~\eqref{length}, we get $k / |\Delta_M| \sim 10^{22} a (k / 100\,\text{Hz}) (10^{-6} \, \text{Gauss} / \bar{B}_0 \sin \theta)$.
Additionally, assuming the radiation-dominated epoch where $\mathcal{H} = H_0 \sqrt{\Omega_{\text{r0}}} / a$, with $H_0 = 100 \, h \,\text{km/sec/Mpc}$ and $\Omega_{\text{r0}} \sim 10^{-5} h^{-2}$ (and $h \sim 0.7$), we find that $k / \mathcal{H} \sim 10^{22} a (k / 100\,\text{Hz})$.
These two results indicate that for a wavenumber in the LIGO band ($k \sim 100 \, \text{Hz}$), Eq.~\eqref{eqsch6} is satisfied when $a \gg 10^{-22}$.
}
\begin{align}
    k \gg |\Delta_M(a)|, \mathcal{H}\,. 
    \label{eqsch6}
\end{align}

Considering the Standard Model (SM) photons as the gauge fields, the onset of graviton-photon conversion should be the decoupling time, $z \sim 1100$. 
Then, even assuming the observational upper limit of the magnetic field magnitude $\bar{B}_0 \sim 10^{-9} ~\text{Gauss}$, the conversion length at the decoupling time becomes $10^3~\text{Gpc}$, far exceeding the present Hubble radius. Therefore, a significant conversion with gravitons is not expected for the SM photons. 
Indeed, the almost perfect blackbody spectrum of CMB indicates that graviton - SM photon conversion is inefficient \cite{Chen:2013gva}. 

On the other hand, the case of dark photons has a different story.
First, the magnitude of the dark magnetic field is observationally less constrained as mentioned in Eq.~\eqref{constraint2}.
Moreover, as long as the dark photons are decoupled from other particles, the onset of the conversion with gravitons can be earlier. 
Then, as seen from Eq.~\eqref{length}, the length scale of the conversion may be shorter. 
Hence, for dark photons, conversion with gravitons can occur efficiently and can leave observational signatures on primordial gravitational waves: This is the case we consider in the rest of this paper.
In the following sections, we study the imprints of the conversion between gravitational waves and dark photons on the intensity and polarization.

Our focus is to study the impact of the conversion on a given magnetic field since the primordial epoch.
Therefore, the analysis in this paper is applicable to any primordial dark magnetic field  produced during inflation  \cite{Turner:1987bw, Ratra:1991bn}.
Note that the strong coupling problem \cite{Demozzi:2009fu} in inflationary magnetogenesis is irrelevant as long as the dark photons are decoupled from the SM.

The expression for the conversion length \eqref{length} implies that conversion can be particularly effective during the radiation-dominated (RD) epoch. Focusing on this epoch, the integrals in Eqs.~\eqref{ysol1} and \eqref{Asol1} can be performed analytically as follows.
From the definition of the conformal time $a\, d\eta = dt$ and the Hubble parameter $H = (da/dt) / a$, it follows that $d\eta = (a^2 H)^{-1} da$. 
Additionally, the Friedmann equation in the RD epoch is given by $H = H_0 \sqrt{\Omega_{\text{r0}} / a^4}$ where $H_0$ is the Hubble constant and $\Omega_{\text{r0}}$ is the present density parameter of radiation.
Thus, we obtain 
\begin{align}
    \eta = \frac{a}{H_0 \sqrt{\Omega_{\text{r0}}}} 
    = 4.6 \times 10^5 \, a ~\text{Mpc} \,, 
    \qquad \text{(RD)}
    \label{eta-a-3}
\end{align}
where we set the origin of the conformal time to be the Big Bang ($a=0$), and we employed the observed values, $H_0 = 100 \, h$ km/sec/Mpc and $\Omega_{\text{r}0}h^2 = 4.2 \times 10^{-5}$ with $h = 0.68$.
Then, the integrals in Eqs.~\eqref{ysol1} and \eqref{Asol1} are explicitly evaluated as 
\begin{align}
    \int_{\eta_{\text{i}}}^\eta d\eta \, \Delta_{M}(a) 
    &= \frac{\kappa}{\sqrt{2}} \frac{\bar{B}_0}{H_0 \sqrt{\Omega_{\text{r0}}}} (\sin \theta) \ln \bigg( \frac{\eta}{\eta_{\text{i}}} \bigg) \,.
    \qquad \text{(RD)}
    \label{intDeltaM}
\end{align}

\section{Conversion in a network of magnetic domains}
\label{sec:network}

\subsection{Modeling of a network of magnetic domains}

Magnetic fields in the universe can have various strengths and coherence lengths at different scales.
For magnetic fields composed of Standard Model photons, observational constraints on the correlation length $s$ have been suggested from magnetohydrodynamic (MHD) turbulence, CMB spectral distortions, gamma-ray observations, and related observations, yielding $s \gtrsim \mathrm{pc}$ (see, e.g., Fig.19 in Ref.~\cite{Durrer:2013pga}).
By contrast, there are currently no observational constraints on the correlation length for magnetic fields composed of dark photons, which are the main focus of this work.
From a theoretical perspective, the finite correlation length of dark magnetic fields may be set as an initial condition at their generation. 
For example, in scenarios where magnetic fields are generated during inflation \cite{Turner:1987bw, Ratra:1991bn} or phase transitions \cite{Vachaspati:1991nm}, the correlation length is determined by the Hubble radius at the time of generation.
Furthermore, decoherence of dark magnetic fields might occur through mechanisms of MHD turbulence, similar to those for ordinary magnetic fields, if a dark plasma exists that couples to dark photons. 
From these considerations, it is natural to expect that dark magnetic fields may exist with finite correlation lengths.

\begin{figure}[tb]
\centering
\includegraphics[scale=0.58]{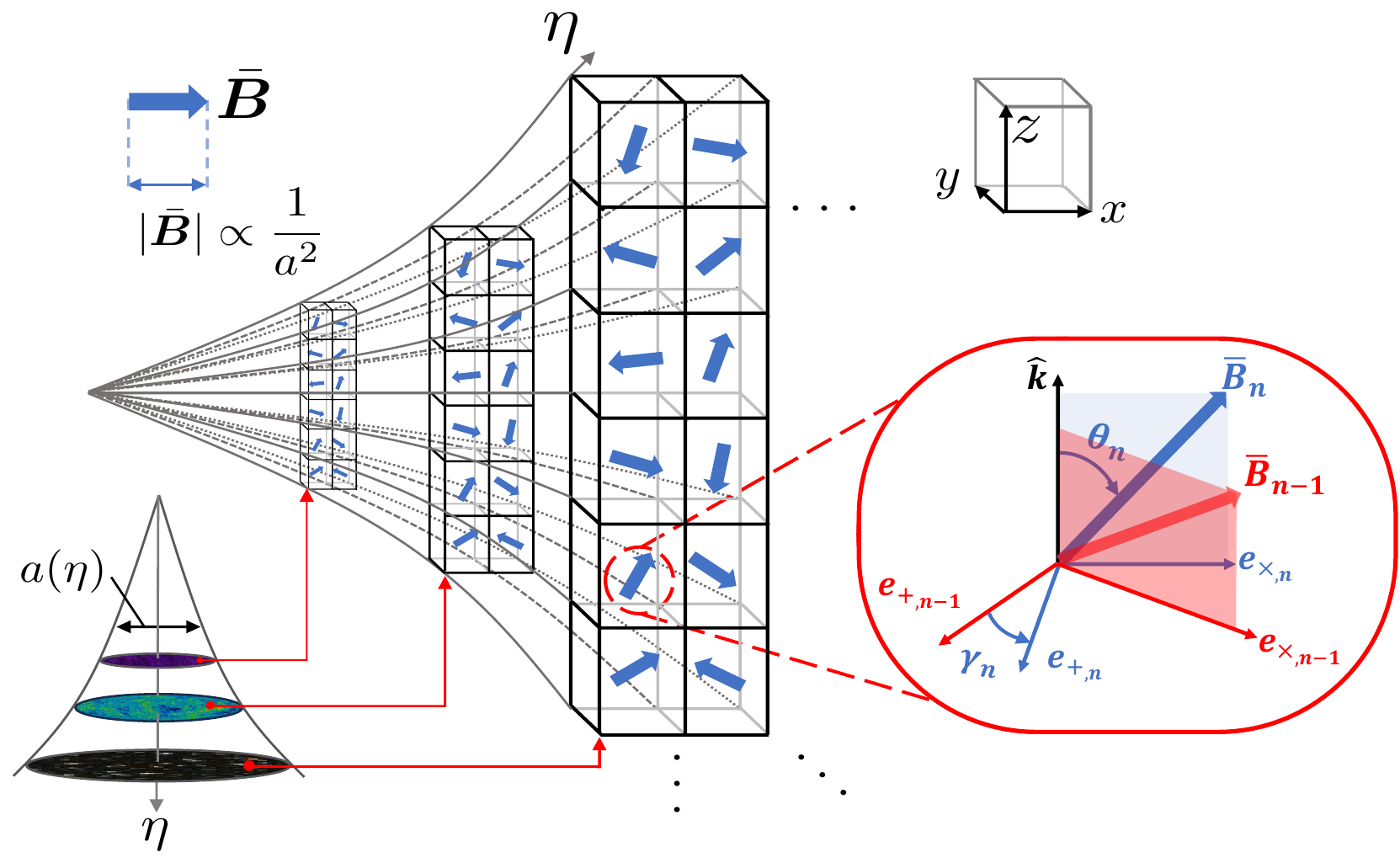}  
\caption{The configuration of the background magnetic fields in our model is illustrated. We assume that gravitational waves and gauge field waves pass through an enormous number of domains with equal comoving sizes. Within each domain, the background magnetic field can be regarded as a uniform vector. The magnitude of the magnetic fields evolves as $|\bar{\bm B}|\propto 1/a^2$.
The index $n$ denotes the sequence number of the magnetic field domains that a gravitational wave passes through, as described in Sec. 3.1. Accordingly, $\bar{\bm{B}}_n$ and $\bar{\bm{B}}_{n-1}$ represent the magnetic field vectors in the $n$-th and $(n-1)$-th domains, respectively.
}
\label{configu}
\end{figure}

To model magnetic fields on the cosmological scale, we construct a network of many coherent magnetic field domains as follows (see Fig.~\ref{configu}).
First, all domains are assumed to have equal comoving size $s$, which compares with coherent scale of the field.  
The background magnetic field is assumed to be uniform within a single domain.
For simplicity, we let the magnitude of the magnetic field $|\bar{\bm{B}}|$ be equal in all domains on a time slice, but allow the direction of the magnetic field to vary from domain to domain. 
Specifically, we assume that the direction of the magnetic field is random for each domain.
A gravitational wave passes through a network of domains with a fixed wavenumber vector $\bm{k}$. 
At the $n$-th domain during propagation, we let $\theta_n$ denote the angle between the wavenumber vector $\bm{k}$ and the magnetic field vector $\bar{\bm{B}}$. 
When moving from the $(n-1)$-th domain into the $n$-th domain, the magnetic field vector rotates by an angle $\gamma_n$ in the polarization plane (the two-dimensional plane normal to the wavenumber vector $\bm{k}$).
Note that our model which consists of a large number of magnetic domains has been employed in Ref.~\cite{Grossman:2002by} as a model of magnetic fields with finite coherence lengths.

\subsection{Time evolution of Stokes parameters}
\label{subsec:evolution}

Recall that we have defined the polarization vectors $\bm{e}^+(\hat{\bm{k}})$ and $\bm{e}^\times(\hat{\bm{k}})$ so that the magnetic field vector $\bar{\bm{B}}$ lies in the $\hat{\bm{k}}$-$\bm{e}^\times(\hat{\bm{k}})$ plane.
Since the magnetic field vector changes the direction when the domain changes, the definition of the polarization vector must also change.
Denoting the polarization vectors in the $n$-th domain by $\{ \bm{e}_{n}^+(\hat{\bm{k}}) , \bm{e}_n^\times(\hat{\bm{k}}) \}$, the relationship between adjacent domains is given by 
\begin{align}
    \begin{pmatrix}
        \bm{e}_{n}^+ (\hat{\bm{k}}) \\
        \bm{e}_{n}^\times (\hat{\bm{k}})
    \end{pmatrix}
    =
    \begin{pmatrix}
        \cos \gamma_n   &\sin \gamma_{n} \\
        - \sin \gamma_{n}   &\cos \gamma_n
    \end{pmatrix}
    \begin{pmatrix}
        \bm{e}_{n-1}^+ (\hat{\bm{k}}) \\
        \bm{e}_{n-1}^\times (\hat{\bm{k}})
    \end{pmatrix}\,.
    \label{eqrot1}
\end{align}
This rotation induces the transformation of the polarization tensors as 
\begin{align}
    \begin{pmatrix}
        {e}_{n,ij}^+ (\hat{\bm{k}}) \\
        {e}_{n,ij}^\times (\hat{\bm{k}})
    \end{pmatrix}
    =
    \begin{pmatrix}
        \cos 2\gamma_n   &\sin 2\gamma_{n} \\
        - \sin 2\gamma_{n}   &\cos 2\gamma_n
    \end{pmatrix}
    \begin{pmatrix}
        {e}_{n-1,ij}^+ (\hat{\bm{k}}) \\
        {e}_{n-1,ij}^\times (\hat{\bm{k}})
    \end{pmatrix}\,.
    \label{eqrot2}
\end{align}

Let $A_{P,n}$ and $y_{P,n}$ ($P=+,\times$) denote the polarization components defined by means of the polarization vectors and tensors in the $n$-th domain. 
According to the rotation \eqref{eqrot1} and \eqref{eqrot2}, the relationship between adjacent polarization components  reads 
\begin{align}
    \begin{pmatrix}
        A_{+,n} (\eta) \\
        A_{\times,n} (\eta)
    \end{pmatrix}
    &=
    \begin{pmatrix}
        \cos \gamma_n   &\sin \gamma_{n} \\
        - \sin \gamma_{n}   &\cos \gamma_n
    \end{pmatrix}
    \begin{pmatrix}
        A_{+,n-1} (\eta) \\
        A_{\times,n-1} (\eta)
    \end{pmatrix}\,,
    \label{Arot1}
    \\
    \begin{pmatrix}
        y_{+,n} (\eta) \\
        y_{\times,n} (\eta)
    \end{pmatrix}
    &=
    \begin{pmatrix}
        \cos 2\gamma_n   &\sin 2\gamma_{n} \\
        - \sin 2\gamma_{n}   &\cos 2\gamma_n
    \end{pmatrix}
    \begin{pmatrix}
        y_{+,n-1} (\eta) \\
        y_{\times,n-1} (\eta)
    \end{pmatrix}\,.
    \label{hrot1}
\end{align}
Here, we omitted the wavenumber vector $\bm{k}$ in the argument of $y_{P,n} (\bm{k}, \eta)$ and $A_{P,n} (\bm{k}, \eta)$ for notational simplicity.

Let $\eta_{n-1}$ be the conformal time at the end of the passage through the $(n-1)$-th domain, which also gives the time of entry into the $n$-th domain.
Setting $\eta=\eta_{n-1}$ in Eqs.~\eqref{Arot1} and \eqref{hrot1} specifies the initial condition for the polarization components defined in the $n$-th domain $A_{P,n}(\eta_{n-1})$ and $y_{P,n}(\eta_{n-1})$ in terms of the polarization components defined in the previous domain $A_{P,n-1}(\eta_{n-1})$ and $y_{P,n-1}(\eta_{n-1})$. 
Inserting the initial condition into the general solutions \eqref{ysol1} and \eqref{Asol1}, we obtain the relationship between the polarization components at the end of the $n$-th domain and those at the end of the $(n-1)$-th domain as 
\begin{align}
    \bm{\varphi}_n
    &= \bm{\Phi}_n \cdot \bm{\Gamma}_n
    \cdot \bm{\varphi}_{n-1} \,.
    \label{eqrec1}
\end{align}
Here, we introduced a four-component vector $\bm{\varphi}_n$ as
\begin{align}
    \bm{\varphi}_n 
    \equiv 
    \big(
        \tilde{y}_{+,n}(\eta_n),
        \tilde{y}_{\times,n}(\eta_n),
        \tilde{A}_{+,n}(\eta_n),
        \tilde{A}_{\times,n}(\eta_n)
    \big)^\top .
    \label{statevec}
\end{align}
We also defined
\begin{align}
    \bm{\Phi}_n &\equiv
    \begin{pmatrix}
        \cos  \Phi_n  &0  &-i\sin \Phi_n &0 \\
        0   &\cos  \Phi_n    &0  &-i\sin \Phi_n  \\
        -i\sin \Phi_n   &0  &\cos  \Phi_n    &0 \\
        0   &-i\sin \Phi_n   &0  &\cos  \Phi_n 
    \end{pmatrix} 
    \label{defmatPhi}
\end{align}
with
\begin{align}
    \Phi_n \equiv \int_{\eta_{n-1}}^{\eta_n} d\eta \, \Delta_{M,n}(a) \,,
    \qquad 
    \Delta_{M,n}(a) \equiv \frac{\kappa}{\sqrt{2}} \frac{\bar{B}_0}{a} \sin \theta_n \,,
    \label{defphi}
\end{align}
and 
\begin{align}
    \bm{\Gamma}_n \equiv
    \begin{pmatrix}
        \cos  2 \gamma_n   &\sin 2 \gamma_n   &0  &0 \\
        - \sin 2 \gamma_n  &\cos  2 \gamma_n  &0  &0 \\
        0  &0   &\cos  \gamma_n   &\sin \gamma_n \\
        0  &0   &- \sin \gamma_n  &\cos  \gamma_n 
    \end{pmatrix}. 
    \label{defmatGamma}
\end{align}

To describe polarization, let us introduce Stokes parameters with the vector $\bm{\varphi}$ defined in Eq.~\eqref{statevec} as 
\begin{align}
    \bm{\varphi}\, \bm{\varphi}^\dag
    &\equiv
    \frac{1}{2}
    \begin{pmatrix}
        I_h + Q_h   &U_h - i V_h    &K - i L   &M - i N \\
        U_h + i V_h &I_h - Q_h      &W - i X   &Y - i Z \\
        K + i L     &W + i X        &I_\gamma + Q_\gamma   &U_\gamma - i V_\gamma \\
        M + i N     &Y + i Z        &U_\gamma + i V_\gamma &I_\gamma - Q_\gamma 
    \end{pmatrix} , 
    \label{stokes2}
\end{align}
where the subscripts $h$ and $\gamma$ represent that the quantities are associated with gravitational waves and gauge fields, respectively.
The subscript labeling the number of domains $n$ is omitted here.
Specifically, the intensity of gravitational waves and gauge fields are denoted by $I_h$ and $I_\gamma$, respectively, and those after passing the $n$-th domain are given by
\begin{align}
    I_h(\bm{k}, \eta_n) &= | \tilde{y}_{+,n}(\bm{k}, \eta_n) |^2 + | \tilde{y}_{\times,n}(\bm{k}, \eta_n) |^2 \,, 
    \label{inth}
    \\
    I_\gamma(\bm{k}, \eta_n) &= | \tilde{A}_{+,n}(\bm{k}, \eta_n) |^2 + | \tilde{A}_{\times,n}(\bm{k}, \eta_n) |^2 \,.
    \label{intA}
\end{align}
The variables $V_h$ and $V_\gamma$ denote the degree of circular polarization of gravitational waves and gauge fields, respectively. Those at the end of the $n$-th domain are given by
\begin{align}
    V_h(\bm{k}, \eta_n) &= i \big[ \tilde{y}_{\times,n}^*(\bm{k}, \eta_n) \, \tilde{y}_{+,n}(\bm{k}, \eta_n) - \tilde{y}_{+,n}^*(\bm{k}, \eta_n) \, \tilde{y}_{\times,n}(\bm{k}, \eta_n) \big]\,, 
    \label{Vh}
    \\
    V_\gamma(\bm{k}, \eta_n) &= i \big[ \tilde{A}_{\times,n}^*(\bm{k}, \eta_n) \, \tilde{A}_{+,n}(\bm{k}, \eta_n) - \tilde{A}_{+,n}^*(\bm{k}, \eta_n) \,\tilde{A}_{\times,n}(\bm{k}, \eta_n) \big] \,. 
    \label{Vgamma}
\end{align}
The variables  $Q_{h/\gamma}$ and $U_{h/\gamma}$ denote the degree of linear polarization of gravitational waves/gauge fields,
\begin{align}
    Q_h(\bm{k}, \eta_n) &= | \tilde{y}_{+,n}(\bm{k}, \eta_n) |^2 - | \tilde{y}_{\times,n}(\bm{k}, \eta_n) |^2 \,, 
    \label{Qh}
    \\
    Q_\gamma(\bm{k}, \eta_n) &= | \tilde{A}_{+,n}(\bm{k}, \eta_n) |^2 - | \tilde{A}_{\times,n}(\bm{k}, \eta_n) |^2 \,,
    \label{Qgamma}
\end{align}
and
\begin{align}
    U_h(\bm{k}, \eta_n) &=  \tilde{y}_{\times,n}^*(\bm{k}, \eta_n) \, \tilde{y}_{+,n}(\bm{k}, \eta_n) + \tilde{y}_{+,n}^*(\bm{k}, \eta_n) \, \tilde{y}_{\times,n}(\bm{k}, \eta_n) \,,  
    \label{Uh}
    \\
    U_\gamma(\bm{k}, \eta_n) &=  \tilde{A}_{\times,n}^*(\bm{k}, \eta_n) \, \tilde{A}_{+,n}(\bm{k}, \eta_n) + \tilde{A}_{+,n}^*(\bm{k}, \eta_n) \,\tilde{A}_{\times,n}(\bm{k}, \eta_n) \,. 
    \label{Ugamma}
\end{align}
Note that $I_{h/\gamma}$ and $V_{h/\gamma}$ are coordinate-independent notions in the sense that they are invariant under rotation of the coordinate axes in the polarization plane, while $Q_{h/\gamma}$ and $U_{h/\gamma}$ are dependent.
In Eq.~\eqref{stokes2}, we also introduced real parameters $K$, $L$, $M$, $N$, $W$, $X$, $Y$, and $Z$, which consist of cross terms of $\tilde{y}_{P}$ and $\tilde{A}_P$ as follows:
\begin{align}
    K(\bm{k}, \eta_n) &=  \tilde{y}_{+,n}(\bm{k}, \eta_n) \, \tilde{A}_{+,n}^*(\bm{k}, \eta_n) + \tilde{y}_{+,n}^*(\bm{k}, \eta_n) \, \tilde{A}_{+,n}(\bm{k}, \eta_n) \,,  
    \label{defK}
    \\
    L(\bm{k}, \eta_n) &= i \big[\tilde{y}_{+,n}(\bm{k}, \eta_n) \, \tilde{A}_{+,n}^*(\bm{k}, \eta_n) - \tilde{y}_{+,n}^*(\bm{k}, \eta_n) \, \tilde{A}_{+,n}(\bm{k}, \eta_n)\big]\,, 
    \label{defL}
    \\
   M(\bm{k}, \eta_n) &=  \tilde{y}_{+,n}(\bm{k}, \eta_n) \, \tilde{A}_{\times,n}^*(\bm{k}, \eta_n) + \tilde{y}_{+,n}^*(\bm{k}, \eta_n) \, \tilde{A}_{\times,n}(\bm{k}, \eta_n) \,, 
   \label{defM}
    \\
   N(\bm{k}, \eta_n) &= i \big[\tilde{y}_{+,n}(\bm{k}, \eta_n) \, \tilde{A}_{\times,n}^*(\bm{k}, \eta_n) - \tilde{y}_{+,n}^*(\bm{k}, \eta_n) \, \tilde{A}_{\times,n}(\bm{k}, \eta_n) \big]\,, 
   \label{defN}
    \\
   W(\bm{k}, \eta_n) &=  \tilde{y}_{\times,n}(\bm{k}, \eta_n) \, \tilde{A}_{+,n}^*(\bm{k}, \eta_n) + \tilde{y}_{\times,n}^*(\bm{k}, \eta_n) \, \tilde{A}_{+,n}(\bm{k}, \eta_n) \,, 
   \label{defW}
    \\
   X(\bm{k}, \eta_n) &= i \big[\tilde{y}_{\times,n}(\bm{k}, \eta_n) \, \tilde{A}_{+,n}^*(\bm{k}, \eta_n) - \tilde{y}_{\times,n}^*(\bm{k}, \eta_n) \, \tilde{A}_{+,n}(\bm{k}, \eta_n)\big]\,, 
   \label{defX}
    \\
   Y(\bm{k}, \eta_n) &=  \tilde{y}_{\times,n}(\bm{k}, \eta_n) \, \tilde{A}_{\times,n}^*(\bm{k}, \eta_n) + \tilde{y}_{\times,n}^*(\bm{k}, \eta_n) \, \tilde{A}_{\times,n}(\bm{k}, \eta_n) \,,
   \label{defY}
    \\
    Z(\bm{k}, \eta_n) &= i \big[\tilde{y}_{\times,n}(\bm{k}, \eta_n) \, \tilde{A}_{\times,n}^*(\bm{k}, \eta_n) - \tilde{y}_{\times,n}^*(\bm{k}, \eta_n) \, \tilde{A}_{\times,n}(\bm{k}, \eta_n)\big]\,.
    \label{defZ}
\end{align}

To study the evolution of the Stokes parameters, we symbolize the matrix in Eq.~\eqref{stokes2} by $\bm{\rho}$.
Then, at the end of the $n$-th domain, we obtain the relation
\begin{align}
    \bm{\rho}_n \equiv \bm{\varphi}_n \, \bm{\varphi}_n^\dag 
    = (\bm{\Phi}_n \cdot \bm{\Gamma}_n) \cdot \bm{\rho}_{n-1} \cdot (\bm{\Phi}_n \cdot \bm{\Gamma}_n)^\dag 
    \,,
    \label{denmat}
\end{align}
where we used Eq.~\eqref{eqrec1}.
 The above relation yields recurrence relations between the $n$-th and $(n-1)$-th Stokes parameters. 
It turns out that not all the Stokes parameters are coupled to each other.
Indeed, the recurrence relations are classified into four classes as follows.

\subsubsection{Class I}
First, let us focus on the part concerning the intensity of gravitational waves and gauge fields, $I_h$ and $I_\gamma$.
It is found that the variables $\{ I_h, I_\gamma, L+Z , X-N \}$ constitute a closed set of equations in the recurrence relation \eqref{denmat} as follows:
\begin{align}
    I_h(\eta_n) + I_\gamma(\eta_n) &= I_h(\eta_{n-1}) + I_\gamma(\eta_{n-1}) \,,
    \label{eqrec201}\\
    I_h(\eta_{n}) - I_\gamma(\eta_n) 
    &= \sin 2\Phi_n \big\{ \cos \gamma_n [ L(\eta_{n-1}) + Z(\eta_{n-1}) ] + \sin \gamma_n [ X(\eta_{n-1}) - N(\eta_{n-1}) ] \big\} 
    \notag \\
    &\quad +
    \cos 2 \Phi_n [ I_h(\eta_{n-1}) - I_\gamma(\eta_{n-1}) ] \,,
    \label{eqrec202}\\
    L(\eta_{n}) + Z(\eta_{n}) 
    &= \cos 2 \Phi_n \big\{ \cos \gamma_n [ L(\eta_{n-1}) + Z(\eta_{n-1}) ] + \sin \gamma_n [ X(\eta_{n-1}) - N(\eta_{n-1}) ] \big\} 
    \notag \\
    &\quad - \sin 2 \Phi_n [ I_h(\eta_{n-1}) - I_\gamma(\eta_{n-1}) ] \,,
    \label{eqrec203}\\ 
    X(\eta_{n}) - N(\eta_{n}) 
    &= \cos \gamma_n [ X(\eta_{n-1}) - N(\eta_{n-1}) ] - \sin \gamma_n [ L(\eta_{n-1}) + Z(\eta_{n-1}) ] \,.
    \label{eqrec204}
\end{align}

\subsubsection{Class II}
The circular polarizations of gravitational waves and gauge field waves are determined by the following closed set of equations
for the variables $\{ V_h, V_\gamma, M-W , K+Y\}$:
\begin{align}
    V_h(\eta_n) + V_\gamma(\eta_n) &= V_h(\eta_{n-1}) + V_\gamma(\eta_{n-1}) \,,
    \label{eqrec209}\\
    V_h(\eta_n) - V_\gamma(\eta_n) 
    &= - \sin 2 \Phi_n \big\{ \cos \gamma_n [ M(\eta_{n-1}) - W(\eta_{n-1}) ] + \sin \gamma_n [ K(\eta_{n-1}) + Y(\eta_{n-1}) ] \big\} 
    \notag \\
    &\quad + \cos 2 \Phi_n [ V_h(\eta_{n-1}) - V_\gamma(\eta_{n-1}) ] \,,
    \label{eqrec210}\\
    M(\eta_n) - W(\eta_{n}) 
    &= \cos 2 \Phi_n \big\{ \cos \gamma_n [ M(\eta_{n-1}) - W(\eta_{n-1}) ] + \sin \gamma_n [ K(\eta_{n-1}) + Y(\eta_{n-1}) ] \big\}
    \notag \\
    &\quad + \sin 2 \Phi_n [ V_h(\eta_{n-1}) - V_\gamma(\eta_{n-1}) ] \,,
    \label{eqrec211}\\ 
    K(\eta_{n}) + Y(\eta_{n}) 
    &= \cos \gamma_n [ K(\eta_{n-1}) + Y(\eta_{n-1}) ] - \sin \gamma_n [ M(\eta_{n-1}) - W(\eta_{n-1}) ] \,.
    \label{eqrec212}
\end{align}

\subsubsection{Class III}
The linear polarizations of gravitational waves and gauge field waves are determined by the following closed set of equations
for the variables $\{Q_h+i \, U_h, Q_\gamma +i \, U_\gamma, L-Z , X+N\}$:
\begin{align}
    Q_h(\eta_n) + i \, U_h(\eta_n) 
    &= \cos^2 \Phi_n \, e^{-4i\gamma_n} [ Q_h(\eta_{n-1}) + i \, U_h(\eta_{n-1}) ] 
    + \sin^2 \Phi_n \, e^{-2i\gamma_n} [ Q_\gamma(\eta_{n-1}) + i \, U_\gamma(\eta_{n-1}) ] 
    \notag \\
    &\quad + \frac{1}{2} \sin 2 \Phi_n \,e^{-3i\gamma_n} \big\{ [ L(\eta_{n-1}) - Z(\eta_{n-1}) ] + i\, [ X(\eta_{n-1}) + N(\eta_{n-1}) ] \big\} \,,
    \label{eqrec205}\\
    Q_\gamma(\eta_{n}) + i \, U_\gamma(\eta_{n}) 
    &= \sin^2 \Phi_n \, e^{-4i\gamma_n} [ Q_h(\eta_{n-1}) + i \, U_h(\eta_{n-1}) ]
    + \cos^2 \Phi_n \, e^{-2i\gamma_n} [ Q_\gamma(\eta_{n-1}) + i \, U_\gamma(\eta_{n-1}) ] 
    \notag \\
    &\quad - \frac{1}{2} \sin 2 \Phi_n \, e^{-3i\gamma_n} \big\{ [L(\eta_{n-1}) - Z(\eta_{n-1})] + i\, [ X(\eta_{n-1}) + N(\eta_{n-1}) ] \big\} \,,
    \label{eqrec206}\\
    L(\eta_n) - Z(\eta_n) 
    &= \cos 2 \Phi_n \big\{ \cos 3 \gamma_n [ L(\eta_{n-1}) - Z(\eta_{n-1}) ] + \sin 3 \gamma_n [ X(\eta_{n-1}) + N(\eta_{n-1})  ] \big\}  
    \notag \\
    &\quad 
    - \frac{1}{2} \sin 2 \Phi_n
    \big\{ e^{-4i\gamma_n} [ Q_h(\eta_{n-1}) + i \, U_h(\eta_{n-1}) ] 
    + \text{c.c.} \notag \\
    &\qquad \qquad \qquad \quad
    - e^{-2i\gamma_n} [ Q_\gamma(\eta_{n-1}) + i \, U_\gamma(\eta_{n-1}) ]
    + \text{c.c.} \big\} \,,
    \label{eqrec207}\\
    X(\eta_n) + N(\eta_n) 
    &= \cos 2 \Phi_n \big\{ \cos 3 \gamma_n [ X(\eta_{n-1}) + N(\eta_{n-1}) ] - \sin 3\gamma_n [ L(\eta_{n-1}) - Z(\eta_{n-1}) ] \big\} 
    \notag \\
    &\quad + \frac{1}{2} \sin 2 \Phi_n \big\{ i \, e^{-4i\gamma_n} [ Q_h(\eta_{n-1}) + i\, U_h(\eta_{n-1}) ] 
    + \text{c.c.} 
    \notag \\
     &\qquad \qquad \qquad \quad
    - i \, e^{-2i\gamma_n} [ Q_\gamma(\eta_{n-1}) + i \, U_\gamma(\eta_{n-1}) ] + \text{c.c.} \big\}\,,
    \label{eqrec208}
\end{align} 
where ``c.c.'' represents the complex conjugate of the previous term.

\subsubsection{Class IV}
Additionally, we can find that the variables $\{ M+W, K-Y \}$ form a closed set although it is redundant for the purpose of studying the intensity and  polarization of gravitational waves:
\begin{align}
    M(\eta_n) + W(\eta_n) 
    &= \cos 3 \gamma_n [ M(\eta_{n-1}) + W(\eta_{n-1}) ] - \sin 3 \gamma_n [ K(\eta_{n-1}) - Y(\eta_{n-1}) ] \,, 
    \label{eqrec213}\\ 
    K(\eta_n) - Y(\eta_n) 
    &=\sin 3 \gamma_n [ M(\eta_{n-1}) + W(\eta_{n-1}) ] + \cos 3 \gamma_n [ K(\eta_{n-1}) - Y(\eta_{n-1}) ]  \,. 
    \label{eqrec214}
\end{align}

In summary, we have obtained four classes of recurrence relations.
Note that the intensity $(I_{h/\gamma})$, circular polarization $(V_{h/\gamma})$, and linear polarization $(Q_{h/\gamma}, U_{h/\gamma})$ are separated.
This stems from the separation of equations of $+$ and $\times$ modes.
Although we have focused here on the minimal coupling, when we take into account 
non-minimal cases, the situation would be different.

\section{Imprints on gravitational wave polarizations} 
\label{sec:imprints}

In the previous section, we derived the relationship between the Stokes parameters in adjacent magnetic field domains. 
In this section, we investigate the asymptotic behavior of the Stokes parameters assuming that the direction of the magnetic field is randomly chosen for each domain. 
Since we have seen that intensity, circular polarization, and linear polarization are separated in the recurrence relations, we can study each of these classes separately.

\subsection{Intensity}

First, let us focus on Class I, i.e., Eqs.~\eqref{eqrec201}--\eqref{eqrec204}, to examine the intensity.
Since the magnetic field direction is assumed to be given probabilistically, we study statistical quantities employing an ensemble average. 
Specifically, the purpose below is to find the asymptotic behavior of the mean (expectation value) of the intensity and the variance of the intensity.

\subsubsection{Mean of intensity}

Equation \eqref{eqrec201} shows that the sum of $I_h$ and $I_\gamma$ does not change even when the domain changes. This indicates that the total intensity of the gravitational waves and gauge fields is conserved,  
\begin{align}
    I_h(\eta) + I_\gamma(\eta) 
    = I_{h,0} + I_{\gamma,0}\,,
    \label{eqasy4}
\end{align}
where $I_{h,0} = I_{h}(\eta_0)$ and $I_{\gamma,0} = I_\gamma(\eta_0)$ denote the initial intensity given before entering the first domain.

Next, let us focus on Eq.~\eqref{eqrec202}. Here we write it down again:
\begin{align}
    I_h(\eta_{n}) - I_\gamma(\eta_n) 
    &= \sin 2\Phi_n \big\{ \cos \gamma_n [ L(\eta_{n-1}) + Z(\eta_{n-1}) ] + \sin \gamma_n [ X(\eta_{n-1}) - N(\eta_{n-1}) ] \big\} 
    \notag \\
    &\quad +
    \cos 2 \Phi_n [ I_h(\eta_{n-1}) - I_\gamma(\eta_{n-1}) ] \,. 
    \label{eqrec101}
\end{align}
As mentioned above, the angles $\theta_n$ and $\gamma_n$ are chosen randomly.
Then, for the purpose of finding the statistical means (expectation values), we can drop the linear terms concerning $\cos \gamma_n$ or $\sin \gamma_n$ in Eq.~\eqref{eqrec101} by averaging over $\gamma_n$. 
On the other hand, the coefficient $\cos 2 \Phi_n$ has the $\theta_n$-dependence as is shown in Eq.~\eqref{defphi}, and the average over $\theta_n$ is nontrivial in general.
Hence, we write the recursion relation for the mean as 
\begin{align}
    \braket{ I_h(\eta_n) - I_\gamma(\eta_n) }
    = \braket{\cos 2 \Phi_n} \cdot \braket{ I_h(\eta_{n-1}) - I_\gamma(\eta_{n-1}) } \,.
    \label{eqrec102}
\end{align}
where the bracket $\braket{\cdots}$ represents the average over various realizations of the magnetic field directions.
It follows from Eq.~\eqref{eqrec102} that 
\begin{align}
    \braket{ I_h(\eta_N) - I_\gamma(\eta_N) } = \Pi_N ( I_{h,0} - I_{\gamma,0} ) \,, 
    \label{eqrec103}
\end{align}
where we defined 
\begin{align}
    \Pi_N \equiv \prod_{n=1}^N \braket{\cos 2\Phi_n} \,. 
    \label{eqasy5}
\end{align}
Combining Eqs.~\eqref{eqasy4} and \eqref{eqrec103}, we obtain
\begin{align}
    \braket{I_h(\eta_N)}
    &= \frac{1+\Pi_N}{2} \, I_{h,0} + \frac{1-\Pi_N}{2} \, I_{\gamma,0} \,,
    \label{eqasy1}
    \\
    \braket{I_\gamma(\eta_N)}
    &= \frac{1+\Pi_N}{2} \, I_{\gamma,0} + \frac{1-\Pi_N}{2} \, I_{h,0} \,.
     \label{eqasy2}
\end{align}

The bracket on the right-hand side in Eq.~\eqref{eqasy5} consists only of the average over the angle $\theta_n$. 
Let us rewrite the definition of $\Phi_n$ in Eq.~\eqref{defphi} as 
\begin{align}
    \Phi_n = \mathcal{M}_n \sin \theta_n \,,
    \label{eqasy6}
\end{align}
where we defined 
\begin{align}
    \mathcal{M}_n \equiv \int_{\eta_{n-1}}^{\eta_n} d\eta \, \frac{\kappa}{\sqrt{2}} \frac{\bar{B}_0}{a} \,.
    \label{eqasy7}
\end{align}
Then, the average over $\theta_n$ yields 
\begin{align}
    \braket{ \cos 2\Phi_n } 
    &\equiv \frac{1}{\pi} \int_0^\pi d\theta_n \, \cos (2\mathcal{M}_n \sin \theta_n) 
    \notag \\
    &= J_0 (2\mathcal{M}_n) \,,
    \label{eqasy8}
\end{align}
where $J_0(\cdots)$ represents the zeroth Bessel function of the first kind.
As derived in Eq.~\eqref{intDeltaM}, in the radiation-dominated epoch, the argument $\mathcal{M}_n$ is evaluated as 
\begin{align}
    \mathcal{M}_n &= \frac{\kappa}{\sqrt{2}} \frac{\bar{B}_0}{H_0 \sqrt{\Omega_{\text{r0}}}} 
    \ln \bigg( \frac{\eta_{n}}{\eta_{n-1}} \bigg) 
    = 0.41 \bigg( \frac{\bar{B}_0}{10^{-6}~\text{Gauss}} \bigg) \ln \bigg( \frac{\eta_{n}}{\eta_{n-1}} \bigg) \,.
    \qquad 
    \text{(RD)}
    \label{rad3}
\end{align}
Note that there is a relation $\eta_n = \eta_{n-1} + s$ since we defined the comoving size of each domain as $s$.

Hereafter, we consider the case that the argument $\mathcal{M}_n$ defined in Eq.~\eqref{eqasy7} is large in a single domain, i.e., $\mathcal{M}_n \gg 1$. 
This essentially corresponds to the case that the length scale of conversion $\Delta_M^{-1}$ is much shorter than the domain size so that the conversion occurs efficiently within a domain.
More specifically, considering the radiation-dominated epoch \eqref{rad3}, we see that $\mathcal{M}_n \gg 1$ is realized when $s \gg \eta_{n-1}$, i.e., the value of the conformal time at entry into a domain is sufficiently smaller than the comoving size of the domain.\footnote{
From the expression for the conformal time in the radiation-dominated epoch \eqref{eta-a-3}, the condition $s \gg \eta_{n-1}$ can be written as $s \gg 4.6 \times 10^5 a_{n-1} \, \text{Mpc}$, where $a_{n-1}$ denotes the scale factor at the end of the passage through the $(n-1)$-th domain.
This inequality is satisfied at any time during the radiation-dominated epoch if $s \gg 10^2 \, \mathrm{Mpc}$.
}
The asymptotic behavior of Eq.~\eqref{eqasy8} for large $\mathcal{M}_n$ is given by
\begin{align}
    \braket{\cos 2\Phi_n} = J_0(2\mathcal{M}_n)
    \to \sqrt{ \frac{1}{\pi \mathcal{M}_n} } \cos \bigg( 2\mathcal{M}_n - \frac{\pi}{4} \bigg) \,.
    \label{largeasy1}
\end{align}
Moreover, in the infinite limit $\mathcal{M}_n \to \infty$, the mean $\langle \cos 2 \Phi_n \rangle$ converges to zero and we obtain 
\begin{align}
    \braket{I_h(\eta)} 
    \to \frac{1}{2} ( I_{h,0}+ I_{\gamma,0} )\,,
    \qquad 
    \braket{I_\gamma(\eta)} 
    \to \frac{1}{2} ( I_{h,0}+ I_{\gamma,0} )\,.
    \label{eqasy3}
\end{align}
This result indicates that the equipartition of gravitational waves and gauge fields is realized irrespective of the initial condition in the limit $\mathcal{M}_n \to \infty$.

\subsubsection{Variance of intensity}

Next, we proceed to study the behavior of the statistical variance of the intensity. This is achieved by taking the square of Eqs.~\eqref{eqrec201}--\eqref{eqrec204} and taking the ensemble average, i.e., the average over the magnetic field directions.
From Eq.\,\eqref{eqrec201}, it follows that 
\begin{align}
    \braket{ [ I_h(\eta_n) + I_\gamma(\eta_n) ]^2 } 
    = (I_{h,0} + I_{\gamma,0})^2 \,.
    \label{eqrec301-2}
\end{align}
On the other hand, from Eqs.~\eqref{eqrec202}--\eqref{eqrec204}, we obtain 
\begin{align}
    \begin{pmatrix}
    \braket{ (I_{h,n} - I_{\gamma,n})^2 } \\
    \braket{ (L_n + Z_n)^2 } \\ 
    \braket{ (X_n - N_n)^2 }   
    \end{pmatrix}
    &= \begin{pmatrix}
       \braket{ \cos^2 2 \Phi_n } 
       & \frac{1}{2} \braket{ \sin^2 2 \Phi_n } 
       & \frac{1}{2} \braket{ \sin^2 2 \Phi_n } \\
       \braket{ \sin^2 2 \Phi_n } 
       & \frac{1}{2} \braket{ \cos^2 2 \Phi_n } 
       & \frac{1}{2} \braket{ \cos^2 2 \Phi_n } \\
       0 & 1/2 & 1/2
    \end{pmatrix}
    \begin{pmatrix}
    \braket{ (I_{h,n-1} - I_{\gamma,n-1})^2 } \\
    \braket{ (L_{n-1} + Z_{n-1})^2 } \\ 
    \braket{ (X_{n-1} - N_{n-1})^2 }    
    \end{pmatrix} \,,
    \label{eqrec302}
\end{align}
where the subscripts $n$ and $n-1$ attached to the parameters represent that they are evaluated at time $\eta_n$ and $\eta_{n-1}$, respectively. 
In the right-hand side of Eq.~\eqref{eqrec302}, the average over the angle $\gamma_n$ was already performed and the average over $\theta_n$ is indicated by the bracket $\braket{\cdots}$.

As we have done in Eq.~\eqref{eqasy3}, focusing on the efficient conversion limit, i.e., $\mathcal{M}_n \to \infty$, we can simplify Eq.~\eqref{eqrec302} as 
\begin{align}
    \begin{pmatrix}
    \braket{ (I_{h,n} - I_{\gamma,n})^2 } \\
    \braket{ (L_n + Z_n)^2 } \\ 
    \braket{ (X_n - N_n)^2 }    
    \end{pmatrix}
    &= \begin{pmatrix}
       1/2 & 1/4 & 1/4 \\
       1/2 & 1/4 & 1/4 \\
       0 & 1/2 & 1/2
    \end{pmatrix}
    \begin{pmatrix}
    \braket{ (I_{h,n-1} - I_{\gamma,n-1})^2 } \\
    \braket{ (L_{n-1} + Z_{n-1})^2 } \\ 
    \braket{ (X_{n-1} - N_{n-1})^2 }    
    \end{pmatrix}
    \equiv \bm{W}_1 
    \begin{pmatrix}
    \braket{ (I_{h,n-1} - I_{\gamma,n-1})^2 } \\
    \braket{ (L_{n-1} + Z_{n-1})^2 } \\ 
    \braket{ (X_{n-1} - N_{n-1})^2 }
    \end{pmatrix} \,. 
    \label{eqrec302-2}
\end{align}
The coefficient matrix $\bm{W}_1$ can be diagonalized as 
\begin{align}
    \bm{P}_1^{-1} \bm{W}_1 \bm{P}_1 
    = \begin{pmatrix}
        1 & 0 & 0 \\
        0 & 1/4 & 0 \\
        0 & 0 & 0
    \end{pmatrix} \,,
    \qquad 
    \bm{P}_1 = \begin{pmatrix}
        1 & -1/2 & 0 \\
        1 & -1/2 & -1 \\
        1 & 1 & 1
    \end{pmatrix}\,.
    \label{eqrec302-3}
\end{align}
Therefore, the asymptotic values can be derived as 
\begin{align}
    \begin{pmatrix}
    \braket{ (I_{h,n} - I_{\gamma,n})^2 } \\
    \braket{ (L_n + Z_n)^2 } \\ 
    \braket{ (X_n - N_n)^2 }   
    \end{pmatrix}
    &= \bm{P}_1 
    \begin{pmatrix}
        1 & 0 & 0 \\
        0 & 1/4 & 0 \\
        0 & 0 & 0
    \end{pmatrix}^n 
    \bm{P}_1^{-1} 
    \begin{pmatrix}
    (I_{h,0} - I_{\gamma,0})^2\\
    (L_0 + Z_0)^2\\ 
    (X_0 - N_0)^2 
    \end{pmatrix}  
    \notag \\
    &\xrightarrow{n \to \infty}
    \bm{P}_1 
    \begin{pmatrix}
        1 & 0 & 0 \\
        0 & 0 & 0 \\
        0 & 0 & 0
    \end{pmatrix}
    \bm{P}_1^{-1} 
    \begin{pmatrix}
    (I_{h,0} - I_{\gamma,0})^2\\
    (L_0 + Z_0)^2\\ 
    (X_0 - N_0)^2 
    \end{pmatrix}  
    \notag \\
    &= \begin{pmatrix}
        1/3 & 1/3 & 1/3 \\
        1/3 & 1/3 & 1/3 \\
        1/3 & 1/3 & 1/3 
    \end{pmatrix}
    \begin{pmatrix}
    (I_{h,0} - I_{\gamma,0})^2\\
    (L_0 + Z_0)^2\\ 
    (X_0 - N_0)^2 
    \end{pmatrix}  \,.
    \label{eqrec302-4}
\end{align}
Combining Eqs.~\eqref{eqrec301-2} and \eqref{eqrec302-4} yields the asymptotic behavior,
\begin{align}
    \braket{ I_h^2(\eta) } + \braket{ I_\gamma^2(\eta) } 
    \to \frac{2}{3} 
    ( I_{h,0}^2 + I_{\gamma,0}^2 + I_{h,0} I_{\gamma,0} )
    + \frac{1}{6} [ (L_0 + Z_0)^2 + (X_0 - N_0)^2 ]\,. 
    \label{eqrec302-5}
\end{align}
Furthermore, it is legitimate to expect equality between $\braket{I_h^2}$ and $\braket{I_\gamma^2}$ in the case of 
efficient conversion. 
In this case, we obtain 
\begin{align}
    \braket{I_h^2(\eta)} \to 
    \frac{1}{3} 
    ( I_{h,0}^2 + I_{\gamma,0}^2 + I_{h,0} I_{\gamma,0} )
    + \frac{1}{12} [ (L_0 + Z_0)^2 + (X_0 - N_0)^2 ] \,.
    \label{eqrec302-6}
\end{align}
This equation gives the asymptotic value of the statistical variance of the gravitational wave intensity in terms of the initial Stokes parameters in the case of efficient conversion through a number of random magnetic field domains.

As a by-product, we can also derive the asymptotic value of the correlation $\braket{I_h I_\gamma}$ in the same case as  
\begin{align}
    \braket{I_h(\eta) \, I_\gamma(\eta)}
    \to \frac{1}{6} ( I_{h,0}^2 + I_{\gamma,0}^2) 
    + \frac{2}{3} I_{h,0} I_{\gamma,0} 
    - \frac{1}{12} [ (L_0 + Z_0)^2 + (X_0 - N_0)^2 ]\,. 
    \label{eqrec302-7}
\end{align}

\subsection{Circular polarization}

Next, we turn to Class II, i.e., Eqs.~\eqref{eqrec209}--\eqref{eqrec212}, to reveal the asymptotic behavior of the circular polarization denoted by $V_{h/\gamma}$. The analysis is parallel to that of intensity.

\subsubsection{Mean of circular polarization}

Equation \eqref{eqrec209} shows that, similarly to intensity, the total circular polarization of the gravitational waves and gauge fields is conserved,   
\begin{align}
    V_h(\eta) + V_\gamma(\eta) 
    = V_{h,0} + V_{\gamma,0} \,,
    \label{eqasy21}
\end{align}
where $V_{h,0} = V_{h}(\eta_0)$ and $V_{\gamma,0} = V_\gamma(\eta_0)$ denote the initial circular polarization.
The conservation of the total circular polarization follows from the fact that the $+$ and $\times$ modes are separated and evolve in the same manner.

As performed in Eq.~\eqref{eqrec102}, we take the average of Eq.~\eqref{eqrec210} over the magnetic field directions and write the recurrence relation for the mean as 
\begin{align}
    \braket{ V_h(\eta_{n}) - V_\gamma(\eta_{n}) } = \braket{ \cos 2 \Phi_n } \cdot  \braket{ V_h(\eta_{n-1}) - V_\gamma(\eta_{n-1}) }  \,. 
    \label{eqrec221}
\end{align}
Comparing Eqs.~\eqref{eqrec102} and \eqref{eqrec221}, we see that the mean of the difference of the circular polarization $\braket{ V_h - V_\gamma }$ has the same structure as that of the intensity $\braket{ I_h - I_\gamma }$.
Combining this with Eq.~\eqref{eqasy21}, the means of the circular polarization are given by 
\begin{align}
    \braket{V_h(\eta_N)}
    &= \frac{1+\Pi_N}{2} \, V_{h,0} + \frac{1-\Pi_N}{2} \, V_{\gamma,0} \,,
    \label{eqasyVh}
    \\
    \braket{V_\gamma(\eta_N)}
    &= \frac{1+\Pi_N}{2} \, V_{\gamma,0}+ \frac{1-\Pi_N}{2} \, V_{h,0}\,,
     \label{eqasyVA}
\end{align}
where $\Pi_N$ is defined in Eq.~\eqref{eqasy5}.
Furthermore, in the efficient conversion limit $\mathcal{M}_n \to \infty$, the degree of circular polarization becomes equally distributed between gravitational waves and gauge fields, 
\begin{align}
    \braket{V_h(\eta)} 
    \to \frac{1}{2} ( V_{h,0}+ V_{\gamma,0} )\,,
    \qquad 
    \braket{V_\gamma(\eta)} 
    \to \frac{1}{2} ( V_{h,0}+ V_{\gamma,0} )\,.
    \label{eqasyV3}
\end{align}

\subsubsection{Variance of circular polarization}

The statistical variance of the circular polarization can be found by taking the square of Eqs.~\eqref{eqrec209}--\eqref{eqrec212}.
First, Eq.\,\eqref{eqrec209} tells us
\begin{align}
    \braket{ [ V_h(\eta_n) + V_\gamma(\eta_n) ]^2 }
    = (V_{h,0} + V_{\gamma,0})^2 \,.
    \label{eqrec309-2}
\end{align}
On the other hand, taking the square of Eqs.~\eqref{eqrec210}--\eqref{eqrec212} and averaging them over the angles $\theta_n$ and $\gamma_n$, we obtain
\begin{align}
    \begin{pmatrix}
        \braket{ (V_{h,n} - V_{\gamma,n})^2 } \\
        \braket{ (M_n - W_n)^2 } \\ 
        \braket{ (K_n + Y_n)^2 } 
    \end{pmatrix} 
    &= \begin{pmatrix}
        \braket{ \cos^2 2\Phi_n } & \frac{1}{2} \braket{ \sin^2 2 \Phi_n } & \frac{1}{2} \braket{ \sin^2 2 \Phi_n } \\
        \braket{ \sin^2 2 \Phi_n }  & \frac{1}{2} \braket{ \cos^2 2 \Phi_n } & \frac{1}{2} \braket{ \cos^2 2 \Phi_n } \\ 
        0 & 1/2 & 1/2
    \end{pmatrix} 
    \begin{pmatrix}
        \braket{ (V_{h,n-1} - V_{\gamma,n-1})^2 } \\
        \braket{ (M_{n-1} - W_{n-1})^2 } \\ 
        \braket{ (K_{n-1} + Y_{n-1})^2 }
    \end{pmatrix} \,.
    \label{eqrec310}
\end{align}
When focusing on the efficient conversion limit $\mathcal{M}_n \to \infty$, we can simplify Eq.~\eqref{eqrec310} as 
\begin{align}
    \begin{pmatrix}
        \braket{ (V_{h,n} - V_{\gamma,n})^2 } \\
        \braket{ (M_n - W_n)^2 } \\ 
        \braket{ (K_n + Y_n)^2 } 
    \end{pmatrix} 
    &= \begin{pmatrix}
        1/2 & 1/4 & 1/4  \\
        1/2 & 1/4 & 1/4  \\ 
        0 & 1/2 & 1/2
    \end{pmatrix} 
    \begin{pmatrix}
        \braket{ (V_{h,n-1} - V_{\gamma,n-1})^2 } \\
        \braket{ (M_{n-1} - W_{n-1})^2 } \\ 
        \braket{ (K_{n-1} + Y_{n-1})^2 } 
    \end{pmatrix}
    \equiv \bm{W}_2 
    \begin{pmatrix}
        \braket{ (V_{h,n-1} - V_{\gamma,n-1})^2 } \\
        \braket{ (M_{n-1} - W_{n-1})^2 } \\ 
        \braket{ (K_{n-1} + Y_{n-1})^2 } 
    \end{pmatrix}\,.
    \label{eqrec310-2}
\end{align}
Note that the coefficient matrix $\bm{W}_2$ is the same as $\bm{W}_1$ in Eq.~\eqref{eqrec302-2}. Therefore, the asymptotic values are obtained in the same way as Eq.~\eqref{eqrec302-4}: 
\begin{align}
    \begin{pmatrix}
        \braket{ (V_{h,n} - V_{\gamma,n})^2 } \\
        \braket{ (M_n - W_n)^2 } \\ 
        \braket{ (K_n + Y_n)^2 } 
    \end{pmatrix} 
    \xrightarrow{n \to \infty} 
    \begin{pmatrix}
        1/3 & 1/3 & 1/3 \\
        1/3 & 1/3 & 1/3 \\
        1/3 & 1/3 & 1/3 
    \end{pmatrix}
    \begin{pmatrix}
        (V_{h,0} - V_{\gamma,0})^2 \\
        (M_0 - W_0)^2 \\ 
        (K_0 + Y_0)^2
    \end{pmatrix} \,. 
    \label{eqrec310-4}
\end{align}
Combining Eqs.~\eqref{eqrec309-2} and \eqref{eqrec310-4}, we obtain the asymptotic behavior,
\begin{align}
    \braket{ V_h^2(\eta) } + \braket{ V_\gamma^2(\eta) } 
    \to 
    \frac{2}{3} ( V_{h,0}^2 + V_{\gamma,0}^2 + V_{h,0} V_{\gamma,0}) + 
    \frac{1}{6} [ (M_0 - W_0)^2 + (K_0 + Y_0)^2 ]\,.
    \label{eqrec310-5}
\end{align}
Furthermore, assuming equality $\braket{V_h^2} = \braket{V_\gamma^2}$, we obtain 
\begin{align}
    \braket{ V_h^2(\eta) } 
    \to 
    \frac{1}{3} ( V_{h,0}^2 + V_{\gamma,0}^2 + V_{h,0} V_{\gamma,0}) + 
    \frac{1}{12} [ (M_0 - W_0)^2 + (K_0 + Y_0)^2 ]\,. 
    \label{eqrec310-6}
\end{align}
This equation represents the asymptotic value of the statistical variance of the circular polarization of gravitational waves in terms of the initial Stokes parameters.
The assumptions in derivation were that conversion is efficient in a single domain and that the waves pass through many domains where the direction of the magnetic field is random.

We can also obtain the asymptotic value of the correlation $\braket{V_h V_\gamma}$ as
\begin{align}
    \braket{V_h(\eta) \, V_\gamma(\eta) } \to 
    \frac{1}{6} (V_{h,0}^2 + V_{\gamma,0}^2) + \frac{2}{3} V_{h,0} V_{\gamma,0} - \frac{1}{12} [ (M_0 - W_0)^2 + (K_0 + Y_0)^2 ]\,. 
    \label{eqrec310-7}
\end{align}

\subsection{Linear polarization}

Finally, we study the asymptotic behavior of the linear polarization denoted by $Q_{h/\gamma}$ and $U_{h/\gamma}$ for which Class III, i.e., the set of Eqs.~\eqref{eqrec205}--\eqref{eqrec208}, is relevant.

\subsubsection{Mean of linear polarization}

As we have done for the intensity $I_{h/\gamma}$ and circular polarization $V_{h/\gamma}$, we can express the parameters $Q_{h/\gamma}$ and $U_{h/\gamma}$ in the $n$-th domain in terms of the Stokes parameters in the $(n-1)$-th domain from Eqs.~\eqref{eqrec205} and \eqref{eqrec206}.
In this case, only linear terms with respect to $\cos \gamma_n$, $\cos 2\gamma_n$, and so on, appear.
Then, averaging over the angle $\gamma_n$ leads to 
\begin{align}
    \braket{Q_h(\eta)} = 
    \braket{Q_\gamma(\eta)} = 
    \braket{U_h(\eta)} =
    \braket{U_\gamma(\eta)} =
    0\,.
    \label{eqasyQU}
\end{align} 
Hence, the means (expectation values) of linear polarization vanish.
Note that the vanishing mean of the linear polarization is a direct consequence of the isotropic randomness of the magnetic field directions. If the linear polarization had a nonzero mean, it would, by definition, imply the existence of a preferred direction in the polarization of gravitational waves.
In our model, however, the magnetic field directions are assumed to be random and statistically isotropic, with no preferred direction.
This randomness erases any preferred direction in the gravitational wave polarization through successive conversion processes.

\subsubsection{Variance of linear polarization}

The statistical variance of linear polarization can be studied by taking the square of Eqs.~\eqref{eqrec205}--\eqref{eqrec208} and taking the average over the magnetic field directions characterized by $\theta_n$ and $\gamma_n$.
Indeed, we obtain the recurrence relation for the average of the square as
\begin{align}
    \begin{pmatrix}
        \braket{ Q_{h,n}^2 + U_{h,n}^2 }\\
        \braket{ Q_{\gamma,n}^2 + U_{\gamma,n}^2 } \\
        \braket{ ( L_n - Z_n )^2 } \\
        \braket{ ( X_n + N_n )^2 }
    \end{pmatrix}   
    &= \begin{pmatrix}
        \braket{ \cos^4 \Phi_n } 
        & \braket{ \sin^4 \Phi_n } 
        & \frac{1}{4} \braket{ \sin^2 2 \Phi_n } 
        &  \frac{1}{4} \braket{ \sin^2 2 \Phi_n } \\
        \braket{ \sin^4 \Phi_n } 
        & \braket{ \cos^4 \Phi_n } 
        & \frac{1}{4} \braket{ \sin^2 2 \Phi_n } 
        &  \frac{1}{4} \braket{ \sin^2 2 \Phi_n } \\
        \frac{1}{2} \braket{ \sin^2 2\Phi_n } 
        & \frac{1}{2} \braket{ \sin^2 2\Phi_n } 
        & \frac{1}{2} \braket{ \cos^2 2\Phi_n } 
        & \frac{1}{2} \braket{ \cos^2 2\Phi_n } 
        \\
        \frac{1}{2} \braket{ \sin^2 2 \Phi_n }
        & \frac{1}{2} \braket{ \sin^2 2 \Phi_n }
        & \frac{1}{2} \braket{ \cos^2 2 \Phi_n }
        & \frac{1}{2} \braket{ \cos^2 2 \Phi_n }
    \end{pmatrix}
    \begin{pmatrix}
        \braket{ Q_{h,n-1}^2 + U_{h,n-1}^2 } \\
        \braket{ Q_{\gamma,n-1}^2 + U_{\gamma,n-1}^2 } \\
        \braket{ ( L_{n-1} - Z_{n-1} )^2 } \\
        \braket{ ( X_{n-1} + N_{n-1} )^2 }
    \end{pmatrix} \,.
    \label{eqrec313}
\end{align}
When focusing on the efficient conversion limit $\mathcal{M}_n \to \infty$, Eq.~\eqref{eqrec313} is further simplified:
\begin{align}
    \begin{pmatrix}
        \braket{ Q_{h,n}^2 + U_{h,n}^2 } \\
        \braket{ Q_{\gamma,n}^2 + U_{\gamma,n}^2 } \\
        \braket{ ( L_n - Z_n )^2 } \\
        \braket{ ( X_n + N_n )^2 } 
    \end{pmatrix}   
    &= \begin{pmatrix}
        3/8 & 3/8 & 1/8 & 1/8 \\
        3/8 & 3/8 & 1/8 & 1/8 \\
        1/4 & 1/4 & 1/4 & 1/4 \\
        1/4 & 1/4 & 1/4 & 1/4
    \end{pmatrix}
    \begin{pmatrix}
        \braket{ Q_{h,n-1}^2 + U_{h,n-1}^2 } \\
        \braket{ Q_{\gamma,n-1}^2 + U_{\gamma,n-1}^2 } \\
        \braket{ ( L_{n-1} - Z_{n-1} )^2 } \\
        \braket{ ( X_{n-1} + N_{n-1} )^2 }
    \end{pmatrix} 
    \equiv \bm{W}_3 
    \begin{pmatrix}
        \braket{ Q_{h,n-1}^2 + U_{h,n-1}^2 } \\
        \braket{ Q_{\gamma,n-1}^2 + U_{\gamma,n-1}^2 } \\
        \braket{ ( L_{n-1} - Z_{n-1} )^2 } \\
        \braket{ ( X_{n-1} + N_{n-1} )^2 }
    \end{pmatrix} \,.
    \label{eqrec313-2}
\end{align}
The coefficient matrix $\bm{W}_3$ can be diagonalized as  
\begin{align}
    \bm{P}_3^{-1} \bm{W}_3 \bm{P}_3 
    &= \begin{pmatrix}
        1 & 0 & 0 & 0 \\
        0 & 1/4 & 0 & 0 \\
        0 & 0 & 0 & 0 \\
        0 & 0 & 0 & 0
    \end{pmatrix} \,, 
    \qquad 
    \bm{P}_3
    = \begin{pmatrix}
        1 & -1/2 & 0 & -1 \\
        1 & -1/2 & 0 & 1 \\
        1 & 1 & -1 & 0 \\
        1 & 1 & 1 & 0 \\
    \end{pmatrix} \,.
    \label{eqrec313-3}
\end{align}
Hence, we find the asymptotic values through the infinite number of domains $n \to \infty$ as 
\begin{align}
    \begin{pmatrix}
        \braket{ Q_{h,n}^2 + U_{h,n}^2 } \\
        \braket{ Q_{\gamma,n}^2 + U_{\gamma,n}^2 } \\
        \braket{ ( L_n - Z_n )^2 } \\
        \braket{ ( X_n + N_n )^2 }
    \end{pmatrix}   
    &= 
    \bm{P}_3 \begin{pmatrix}
        1 & 0 & 0 & 0 \\
        0 & 1/4 & 0 & 0 \\
        0 & 0 & 0 & 0 \\
        0 & 0 & 0 & 0
    \end{pmatrix}^n 
    \bm{P}_3^{-1} 
    \begin{pmatrix}
        Q_{h,0}^2 + U_{h,0}^2\\
        Q_{\gamma,0}^2 + U_{\gamma,0}^2  \\
        ( L_0 - Z_0 )^2 \\
        ( X_0 + N_0 )^2
    \end{pmatrix} 
    \notag \\
    &\xrightarrow{n \to \infty} 
    \bm{P}_3
    \begin{pmatrix}
        1 & 0 & 0 & 0 \\
        0 & 0 & 0 & 0 \\
        0 & 0 & 0 & 0 \\
        0 & 0 & 0 & 0
    \end{pmatrix}
    \bm{P}_3^{-1} 
    \begin{pmatrix}
        Q_{h,0}^2 + U_{h,0}^2\\
        Q_{\gamma,0}^2 + U_{\gamma,0}^2  \\
        ( L_0 - Z_0 )^2 \\
        ( X_0 + N_0 )^2
    \end{pmatrix} 
    \notag \\
    &= \begin{pmatrix}
        1/3 & 1/3 & 1/6 & 1/6 \\
        1/3 & 1/3 & 1/6 & 1/6 \\
        1/3 & 1/3 & 1/6 & 1/6 \\
        1/3 & 1/3 & 1/6 & 1/6
    \end{pmatrix}
    \begin{pmatrix}
        Q_{h,0}^2 + U_{h,0}^2\\
        Q_{\gamma,0}^2 + U_{\gamma,0}^2  \\
        ( L_0 - Z_0 )^2 \\
        ( X_0 + N_0 )^2
    \end{pmatrix}  \,.
    \label{eqrec313-4}
\end{align}
In particular, the first component represents the asymptotic value of the variance of the linear polarization of gravitational waves in terms of the initial Stokes parameters:
\begin{align}
    \braket{ Q_h^2(\eta) } + \braket{ U_h^2(\eta) } 
    \to 
    \frac{1}{3} [( Q_{h,0}^2 + U_{h,0}^2) + ( Q_{\gamma,0}^2 + U_{\gamma,0}^2)]
    +  \frac{1}{6} [ (L_0 - Z_0)^2 + (X_0 + N_0)^2 ]\,.
    \label{eqrec314-1}
\end{align}

\section{Discussion}
\label{sec:discussion0}

In the previous section, we obtained the asymptotic values of the intensity and polarization of gravitational waves and dark photons affected by the sequence of the graviton - dark photon conversion. 
In this section, we summarize the assumptions we have used to obtain those results, and briefly discuss the observational feasibility of the signals.

\subsection{Summary of assumptions}

Throughout this paper, we have assumed that direct coupling between dark photons and other matter is completely negligible.
For primordial magnetic fields composed of such dark photons, the only observational constraint comes from Big Bang Nucleosynthesis as mentioned in Eq.~(2.12). 
We have considered magnetic fields that satisfy this observational constraint.

Regarding the coupling between gravitons and dark photons,  only minimal coupling was assumed as in Eq.~(2.1). 
This is because it is theoretically the most robust in the sense that in this framework the Planck mass is the only parameter and no unknown parameters are contained. 

We assumed that the magnetic fields have a coherence length beyond which  the directions are uncorrelated and randomly given.
We are interested in situations where the conversion of gravitational waves and dark photons is efficient.    This situation can be easily realized because the magnetic field strength increases as we go back in time. In this situation, the conversion is saturated, so the exact values of the magnetic field strength and coherence length are no longer relevant to the result.

Finally, the dark photons were assumed to be massless in this paper. 
We leave the analysis of the massive case for future work.
A possible effect of a finite mass might be that the conversion rate becomes frequency dependent as discussed e.g.~in Ref.~\cite{Deffayet:2001pc}.

\subsection{Observational feasibility}

In the previous section, we derived the formulas for the asymptotic values of the intensity and polarizations in terms of given initial values.
The initial values would be provided by some other primordial scenarios of one's interest.

As a simple example, let us consider an initial condition in which gravitational waves and dark photons are initially present in roughly equal amounts ($I_{h,0} \sim I_{\gamma,0}$), and only dark photons have an initial circular polarization ($V_{h,0} =0,  V_{\gamma,0} \neq 0$). 
    (For example, circularly polarized gauge fields can be generated in a nontrivial axion background \cite{Garretson:1992vt}.)
    In this case, as indicated by Eq.~(4.26), after the efficient conversion, the circular polarization of the dark photons is transferred to that of the gravitational waves as 
    $\langle{V_h}\rangle / \langle{I_h}\rangle \sim V_{\gamma,0}/(2I_{h,0})$.
    In Refs.~\cite{Seto:2007tn, Seto:2008sr}, it is shown that a network of ground-based interferometers, including LIGO, can probe the circular polarization of the gravitational wave background $V_h$ with SNRs on the same order as that for the intensity $I_h$.
    Therefore, if $V_{\gamma,0} \sim I_{h,0}$, the circular polarization of the gravitational waves originating from the dark photons can be detected simultaneously once the intensity
    is observed.

Let us also comment on the observational implications of the variances computed in this work.
The variances indicate that, in each independent gravitational wave observation, the intensity and polarizations fluctuate due to conversion through random magnetic fields.
Here, ``independent gravitational wave observation'' can be interpreted, for example, as observing background primordial gravitational waves in different directions, since independent magnetic field domains are expected to be realized along different lines of sight.
Hence, these variances can, in principle, be tested by performing gravitational wave observations many times (in many different directions) and collecting statistics.
These variances will be observationally important because they quantify the estimation error for the intensity and polarizations of primordial gravitational waves.
For instance, when taking the average of $N$ independent observations to estimate the intensity and polarizations, the estimation error scales as $\sigma/\sqrt{N}$, where $\sigma$ denotes the standard deviation (i.e., the square root of the variance).
Furthermore, observational confirmation of the predicted variance itself would provide evidence for the conversion through random magnetic fields.

    It will be important to study to what extent the results in this paper depend on the setup. In particular, non-minimal coupling between gravitons and dark photons and peculiar magnetic field configurations may alter the results and the observational prospects.
    Additionally, while the massless dark photons considered in this paper yielded frequency-independent results, the massive dark photons would produce a frequency dependence of the signal.
    We leave these issues for future work.

\section{Conclusion}
\label{sec:discussion}

Gravitational waves will offer a novel and powerful tool for exploring the early universe. Furthermore, since gravitational waves have universal coupling with particles in the dark sector, they can be useful in probing the dark sector.
In this paper, we considered the conversion processes between gravitons and dark photons in cosmological dark magnetic fields and investigated their imprints on the gravitational waves.
Specifically, we considered a sequence of magnetic domains in which the direction of the magnetic field is randomly given. Then, we studied the asymptotic behavior of the intensity and polarization of gravitational waves when they pass through many domains.

As indicated in Sec.~\ref{subsec:evolution}, the evolution equations for gravitational wave intensity, circular polarization, and linear polarization are separated in the framework of minimal coupling between gravitons and dark photons. 
This is clearly different from the case of axion-photon conversion \cite{Bassan:2010ya, Masaki:2017aea}, where axions are converted only to the polarization component of photons parallel to the magnetic field.

We explicitly showed that the statistical means (expectation values) of the intensity and circular polarization become equally distributed between gravitational waves and dark photons as Eqs.~\eqref{eqasy3} and \eqref{eqasyV3} when the conversion is efficient.
Besides, the mean of the linear polarization vanishes as Eq.~\eqref{eqasyQU}, which is due to the randomness of the magnetic field directions.
Going further, we obtained asymptotic values of the statistical variance of the intensity and polarization of gravitational waves in terms of the initial Stokes parameters as Eqs.~\eqref{eqrec302-6}, \eqref{eqrec310-6}, and \eqref{eqrec314-1}.
We should emphasize that the obtained results can be used to infer the initial polarization of dark photons. 
For example, if the primordial circular polarization of dark photons exists, it can be transferred to that of gravitational waves as Eq.~\eqref{eqasyVh}. 
Hence, we can get information on primordial dark photons through the observations of gravitational waves.  
Furthermore, observational confirmation of the predicted variance \eqref{eqrec310-6} will corroborate that the conversion occurred through random magnetic fields.

\section*{Acknowledgments}
We appreciate Hidetoshi Omiya and Ann Mukuno for their useful comments and daily discussions of related topics.
K.\ N. was supported by JSPS KAKENHI Grant Number JP24KJ0117.
J.\ S. was in part supported by JSPS KAKENHI Grant Numbers JP23K22491, and JP24K21548.
K.\ U. was supported by JSPS KAKENHI Grant Number JP24K17050.
Z.\ W. appreciates Kobe University for Hospitality when the project is initiated. Z.\ W. was supported from Ministry of Science and Technology of China by a special grant.

\printbibliography
\end{document}